\documentclass[aps,
prd,
onecolumn,
nofootinbib,
superscriptaddress,10pt
]{revtex4-2}
\usepackage{amsmath,amssymb,amsfonts}
\usepackage{bm}
\usepackage{graphicx}
\usepackage{color}
\usepackage[hidelinks]{hyperref}
\begin{document}
\title{Entanglement Harvesting Landscapes: A Common Computational Framework}

\author{Rahul Nigam}\thanks{ \texttt{rahul.nigam@hyderabad.bits-pilani.ac.in}}
\affiliation{Department of Physics, BITS Pilani Hyderabad Campus, Hyderabad 500078, India}

\author{Ujjwala Vadrevu}\thanks{ \texttt{ujjwala.vadrevu@gmail.com}}
\affiliation{Independent researcher}

\author{Akshay Kulkarni}\thanks{ \texttt{p20230072@hyderabad.bits-pilani.ac.in}} 
\affiliation{Department of Physics, BITS Pilani Hyderabad Campus, Hyderabad 500078, India}

\begin{abstract}
Entanglement harvesting provides an operational means of probing the nonlocal correlation structure of quantum fields through localized detector--field interactions. Although harvesting has been investigated in a wide variety of spacetime backgrounds and quantum field states, existing studies largely focus on individual detector configurations or specific physical settings. In this work, we develop a unified perturbative framework for entanglement harvesting with localized Unruh--DeWitt detectors, providing a general formalism applicable to different detector trajectories and quantum field states through the corresponding Wightman function. By expressing the detector response in terms of dimensionless variables and analytically reducing the perturbative integrals to numerically efficient representations, the framework enables a systematic exploration of multidimensional harvesting landscapes and phase boundaries. As representative applications, we investigate inertial detectors in the Minkowski vacuum, thermal field states, and uniformly accelerated detectors undergoing Born-rigid motion. In the latter case, the detectors have different proper accelerations and hence different Unruh temperatures, leading in general to unequal local excitation probabilities. The harvested entanglement is therefore evaluated by accounting for both local detector responses and the nonlocal exchange amplitude. We find that increasing temperature enhances the local detector excitation probability while generally suppressing the magnitude of the nonlocal exchange amplitude. Consequently, within the perturbative Gaussian-switching framework and over the parameter range investigated, thermal effects suppress the harvested entanglement and reduce the spatial extent of the harvesting region. The resulting harvesting landscapes provide a global characterization of detector--field interactions and demonstrate how temperature and observer motion reshape the competition between local detector noise and nonlocal field correlations. The framework developed here provides a unified and readily extensible methodology for investigating entanglement harvesting across a broad class of relativistic quantum systems.
\end{abstract}

\maketitle

\section{Introduction}

Quantum field theory predicts that even the vacuum is far from being an empty state. Instead, it possesses nontrivial quantum correlations that extend across spacetime and arise from the intrinsic structure of the field itself. These correlations are reflected in the vacuum expectation values of field operators and are responsible for a wide range of relativistic quantum phenomena. However, because vacuum correlations are fundamentally nonlocal, they cannot be accessed directly through measurements performed at a single spacetime point. An important question therefore arises: can these field correlations be converted into observable quantum correlations between localized physical systems?

An operational answer to this question was first provided by Reznik, who demonstrated that two spatially separated localized quantum systems interacting locally with a quantum field can become entangled despite having no direct interaction with one another \cite{Reznik2003}. This remarkable result established that vacuum entanglement is not merely a formal property of quantum field theory, but can be extracted and converted into experimentally accessible entanglement between localized probes. Shortly afterwards, it was shown that the harvested correlations are sufficiently strong to violate Bell inequalities, thereby demonstrating that the extracted entanglement possesses genuine nonclassical character \cite{ReznikRetzkerSilman2005}. These pioneering works introduced what is now known as \emph{entanglement harvesting}, which has since become one of the central paradigms of relativistic quantum information.

In the standard harvesting protocol, two spatially separated Unruh--DeWitt detectors couple locally to a quantum field for a finite duration. Although each detector interacts only with the field at its own spacetime location, the final detector state contains information about the nonlocal structure of the underlying quantum field. To leading order in perturbation theory, the detector dynamics are governed entirely by the field two-point correlation function through the corresponding Wightman function. The harvested entanglement results from a competition between two physically distinct processes: local detector excitations generated by vacuum fluctuations, which introduce noise, and nonlocal field correlations, which generate quantum coherence between the detectors. Entanglement harvesting therefore provides a uniquely operational probe of quantum field correlations that cannot be inferred from the response of an isolated detector alone.

During the past two decades, entanglement harvesting has evolved into a broad and rapidly developing research area. Early investigations established the basic mechanism of vacuum entanglement extraction and explored its dependence on detector separation, interaction duration, switching profiles, and detector energy gap \cite{PozasKerstjens2015}. Subsequent work clarified the interplay between local detector noise and nonlocal exchange amplitudes, leading to quantitative harvesting criteria and a detailed understanding of the role played by the detector parameters and the quantum state of the field \cite{Simidzija2018}. The influence of spacetime structure on entanglement harvesting has also been explored in settings with nontrivial global topology. In particular, Henderson \textit{et al.} demonstrated that the topology of an \(\mathbb{RP}^{2}\) geon, compared with its BTZ black-hole counterpart, can significantly affect the entanglement harvested by two localized detectors, with the effect depending on the detector energy gap and black-hole mass \cite{Henderson_Mann_Twist}. At the same time, the detector model itself has been refined considerably. Realistic atom--field interactions have been investigated using hydrogen-like atoms coupled to the electromagnetic vacuum \cite{Ng2018}, while recent developments have extended the formalism to fully relativistic detector models capable of describing harvesting beyond the standard approximations \cite{Perche2024}. Other studies have shown that harvesting may also be influenced by measurements performed on the field itself \cite{MaesoGarcia2023}, and that localized detector networks can be used to reconstruct the underlying entanglement structure of quantum fields \cite{Torres2023}. These developments illustrate that entanglement harvesting is no longer viewed merely as a theoretical curiosity, but rather as an operational framework for probing the structure of relativistic quantum fields.

A parallel line of research has investigated how harvesting is modified by the physical environment in which the detector--field interaction takes place. Since the detector response depends explicitly on the underlying field correlations, changing the quantum state of the field or the detector trajectories naturally modifies the harvested entanglement. Thermal quantum fields provide one of the earliest and most extensively studied examples of this behaviour, where thermal excitations compete with the nonlocal quantum correlations responsible for harvesting \cite{Simidzija2018}. Detector acceleration introduces another important modification through the Unruh effect, leading to acceleration-assisted as well as acceleration-induced suppression of harvesting depending on the detector configuration \cite{SaltonMann2015}. Beyond flat spacetime, harvesting has been investigated in expanding cosmological backgrounds \cite{SteegMenicucci2009}, de Sitter spacetime \cite{Lin2024}, cosmic string geometries \cite{Ji2024}, electromagnetic quantum fields \cite{Lindel2024}, and more recently within conformal field theories \cite{Wurtz2026}. Collectively, these studies have demonstrated that entanglement harvesting provides a sensitive probe of spacetime geometry, observer motion, field state, and the causal structure of relativistic quantum systems.

Despite this remarkable diversity of applications, most existing investigations remain configuration specific. Typically, each study considers a particular spacetime geometry, detector trajectory, or quantum field state and develops an analytical and numerical treatment tailored to that specific situation. While this approach has produced numerous important physical insights, it tends to obscure an important structural feature of the detector formalism. At leading order in perturbation theory, the detector dynamics possess a universal mathematical structure. Once the appropriate Wightman function has been specified, the perturbative evolution, reduced detector state, and harvesting criterion are obtained through exactly the same detector formalism, irrespective of the underlying spacetime or observer motion. From this perspective, inertial observers, accelerated observers, thermal field states, and more general relativistic settings differ only through the corresponding field correlation functions, while the detector dynamics themselves remain unchanged.

A second limitation concerns the exploration of parameter space. Even for the simplest harvesting protocol involving two identical detectors, the harvested entanglement depends simultaneously on detector separation, energy gap, interaction duration, switching profile, detector trajectory, coupling strength, and the quantum state of the field. Existing studies generally investigate this multidimensional parameter space through one- or two-dimensional parameter scans designed to illustrate representative physical behaviour. Although such analyses successfully identify particular harvesting regimes, they provide only a partial view of the global structure of the detector response. From both conceptual and computational perspectives, it is therefore natural to regard entanglement harvesting as a multidimensional \emph{landscape}, whose geometry encodes the accessibility, robustness, and scaling properties of quantum correlations under varying physical conditions. Such a viewpoint not only provides a more complete characterization of detector--field interactions, but also facilitates systematic comparisons across different physical scenarios.

The present work is motivated by these observations. Rather than treating individual harvesting scenarios independently, we develop a unified theoretical and computational framework based on the general perturbative dynamics of localized Unruh--DeWitt detectors. Beginning from a geometry-independent detector formalism, we derive the reduced two-detector density matrix in terms of the Wightman function and construct a common computational pipeline that transforms the detector response into numerically efficient integral representations. Within this framework, the detector model, perturbative expansion, analytical reduction, numerical implementation, and entanglement diagnostics remain identical for every physical scenario considered. The only quantity that changes from one situation to another is the corresponding Wightman function, thereby clearly separating the universal aspects of detector dynamics from the physical information associated with the underlying quantum field and spacetime.

As representative applications of the framework, we investigate entanglement harvesting by inertial detectors in the Minkowski vacuum, detectors interacting with thermal quantum fields, and uniformly accelerated detectors undergoing Born-rigid motion. The last scenario introduces an important distinction: the two detectors have different proper accelerations and hence different Unruh temperatures, leading, in general, to unequal local excitation probabilities. Consequently, the harvested entanglement must be evaluated by accounting for both local detector responses together with the nonlocal exchange amplitude. Together, these examples demonstrate how modifications of the Wightman function arising from temperature and observer motion reshape the harvesting landscape by altering both the local detector responses and the nonlocal field correlations responsible for entanglement extraction. Although these scenarios serve primarily as benchmark applications, the methodology developed here is considerably more general and may be extended to arbitrary detector trajectories, quantum field states, and curved spacetime backgrounds without modification of the underlying computational framework.

The remainder of this paper is organized as follows. Section~II develops the general perturbative formalism describing localized detector--field interactions and derives the reduced detector density matrix in terms of the Wightman function. Section~III introduces the unified computational framework and presents the analytical reduction and numerical methodology employed throughout this work. Section~IV establishes the benchmark harvesting landscape for inertial detectors in the Minkowski vacuum. Sections~V and VI extend the analysis to thermal quantum fields and uniformly accelerated observers, respectively. Finally, we summarize our principal conclusions and discuss possible extensions of the present framework to more general relativistic quantum systems.
\section{General Formalism}

\subsection{Detector--Field Interaction}

We consider two identical localized Unruh--DeWitt (UDW) detectors
interacting with a real scalar quantum field. The UDW detector
provides an effective description of the interaction between a
localized quantum system and a quantum field, capturing the
essential features of light--matter coupling while avoiding
unnecessary microscopic details. Throughout this work, the
detectors are modelled as identical two-level systems with energy
gap $\Omega$, following prescribed worldlines
$x_{\nu}(\tau_\nu)$, where $\nu=A,B$ labels the detectors and
$\tau_\nu$ denotes the corresponding proper time.

To describe the evolution of the combined detector--field system,
we introduce a common time parameter $t$ with respect to which
the interaction is time ordered. The interaction-picture
Hamiltonian is then written as

\begin{equation}
H_I(t)
=
\lambda
\sum_{\nu=A,B}
\frac{d\tau_\nu}{dt}
\chi_\nu(\tau_\nu(t))
\mu_\nu(\tau_\nu(t))
\phi\!\left(x_\nu(\tau_\nu(t))\right),
\label{eq:HI}
\end{equation}

where $\lambda$ is the detector--field coupling constant,
$\chi_\nu(\tau_\nu)$ is the switching function controlling the
interaction duration, $\mu_\nu(\tau_\nu)$ is the detector
monopole operator, and $\phi(x)$ is the scalar field evaluated
along the detector trajectory. The factor $d\tau_\nu/dt$
accounts for the conversion between the detector's proper time
and the common evolution parameter.

For a two-level detector, the monopole operator evolves according
to

\begin{equation}
\mu_\nu(\tau_\nu)
=
\sigma_\nu^{+}e^{i\Omega\tau_\nu}
+
\sigma_\nu^{-}e^{-i\Omega\tau_\nu},
\label{eq:monopole}
\end{equation}

where $\sigma_\nu^{\pm}$ are the raising and lowering operators
acting on the detector Hilbert space. The energy gap $\Omega$
therefore determines the characteristic frequency to which the
detector is sensitive.

To avoid artificial ultraviolet effects associated with sudden
switching, we employ the smooth Gaussian profile

\begin{equation}
\chi_\nu(\tau_\nu)
=
\exp\!\left(
-\frac{\tau_\nu^{2}}
{2\sigma^{2}}
\right),
\label{eq:gaussian}
\end{equation}

where $\sigma$ characterizes the interaction timescale. Gaussian
switching is particularly advantageous because it admits several
analytical simplifications while remaining physically well
motivated.

Initially, the detectors are assumed to be uncorrelated and
prepared in their joint ground state,

\begin{equation}
|g_Ag_B\rangle,
\end{equation}

while the scalar field is prepared in an arbitrary quantum state
$|\Psi\rangle$. The initial state of the combined detector--field
system is therefore

\begin{equation}
\rho_0
=
|g_Ag_B\rangle
\langle g_Ag_B|
\otimes
|\Psi\rangle
\langle\Psi|.
\label{eq:initialstate}
\end{equation}

The formalism developed below applies to arbitrary detector
trajectories and quantum states of the field. Different physical
situations, including inertial and accelerated observers, vacuum
and thermal field states, or more general curved spacetime
backgrounds, are distinguished through the detector trajectories
and the corresponding field correlation functions. As will become
evident in the perturbative treatment, all dependence on the
underlying spacetime geometry and quantum state enters through
the associated Wightman two-point function.

In the following subsection we evaluate the detector evolution perturbatively in the weak-coupling limit and derive the reduced density matrix governing the dynamics of the detector pair.

\subsection{Perturbative Evolution and Reduced Detector State}

Throughout this work we assume that the detector--field interaction is sufficiently weak for the evolution of the combined detector--field system to be treated perturbatively in the coupling constant $\lambda$. In the interaction picture, the evolution operator is given by the Dyson series

\begin{equation}
U
=
\mathcal{T}
\exp
\left[
-i
\int_{-\infty}^{\infty}
d\tau\,
H_I(\tau)
\right],
\label{eq:Dyson}
\end{equation}

where $\mathcal{T}$ denotes time ordering. Expanding Eq.~(\ref{eq:Dyson}) to second order in the coupling yields

\begin{equation}
U=\mathbb{I}+U^{(1)}+U^{(2)}+\mathcal{O}(\lambda^3),
\end{equation}
with
\begin{align}
U^{(1)}
&=
-i
\int_{-\infty}^{\infty}
d\tau\,
H_I(\tau),
\\
U^{(2)}
&=
-
\int_{-\infty}^{\infty}
d\tau
\int_{-\infty}^{\tau}
d\tau'
H_I(\tau)
H_I(\tau').
\end{align}

The evolved density operator of the combined detector--field system is therefore

\begin{equation}
\rho
=
U
\rho_0
U^\dagger,
\end{equation}

where $\rho_0$ is the initial state defined in Eq.~(\ref{eq:initialstate}). Retaining terms up to order $\lambda^2$ gives

\begin{align}
\rho
=
\rho_0
+
U^{(1)}\rho_0
+
\rho_0 U^{(1)\dagger}
+
U^{(2)}\rho_0
+
\rho_0 U^{(2)\dagger}
+
U^{(1)}\rho_0U^{(1)\dagger}
+
\mathcal{O}(\lambda^3).
\label{eq:rhoExpansion}
\end{align}

Since our interest lies in the detector dynamics, the field degrees of freedom are traced out,

\begin{equation}
\rho_{AB}
=
\mathrm{Tr}_{\phi}
\left(
\rho
\right),
\label{eq:ReducedState}
\end{equation}

to obtain the reduced detector state.

For all field states considered in this work, the one-point function vanishes,

\begin{equation}
\langle
\Psi
|
\phi(x)
|
\Psi
\rangle
=
0,
\end{equation}

so that every first-order contribution disappears after tracing over the field. The leading nontrivial corrections to the detector dynamics therefore arise at second order in perturbation theory.

The trace over the field naturally introduces the positive-frequency two-point correlation function,

\begin{equation}
W(x,x')
=
\langle
\Psi
|
\phi(x)
\phi(x')
|
\Psi
\rangle,
\label{eq:Wightman}
\end{equation}

known as the Wightman function. This quantity contains complete information about both the quantum state of the field and the spacetime geometry relevant to the leading-order detector dynamics.

Using Eq.~(\ref{eq:Wightman}), the reduced density matrix of the
detector pair assumes the general form

\begin{equation}
\rho_{AB}
=
\begin{pmatrix}
1-P_A-P_B & 0 & 0 & X\\
0 & P_B & C & 0\\
0 & C^{*} & P_A & 0\\
X^{*} & 0 & 0 & 0
\end{pmatrix}
+
\mathcal{O}(\lambda^{4}),
\label{eq:DensityMatrix}
\end{equation}

written in the ordered basis

\[
\{
|g_Ag_B\rangle,
|g_Ae_B\rangle,
|e_Ag_B\rangle,
|e_Ae_B\rangle
\}.
\]

Here $P_A$ and $P_B$ denote the local excitation probabilities of
detectors A and B, respectively, $X$ is the nonlocal exchange
amplitude responsible for entanglement harvesting, and $C$
represents the coherence between the two single-excitation states.
Although the detectors may have identical internal properties,
their local excitation probabilities need not be equal, since
these also depend on their respective trajectories and switching
functions. In configurations where the two local responses
coincide, we may write $P_A=P_B\equiv P$.

The detector response is expressed directly in terms of the Wightman
function. The local excitation probability of detector $D=A,B$ is

\begin{equation}
P_D
=
\lambda^2
\int
d\tau_D
\int
d\tau_D'
\,
\chi_D(\tau_D)
\chi_D(\tau_D')
e^{-i\Omega(\tau_D-\tau_D')}
W
\!\left(
x_D(\tau_D),
x_D(\tau_D')
\right),
\label{eq:PGeneral}
\end{equation}

while the exchange amplitude is

\begin{equation}
X
=
-\lambda^2
\int
d\tau
\int
d\tau'
\,
\chi_A(\tau)
\chi_B(\tau')
e^{i\Omega(\tau+\tau')}
W
\!\left(
x_A(\tau),
x_B(\tau')
\right).
\label{eq:XGeneral}
\end{equation}

These quantities determine the leading-order dynamics of the
detector pair. The local responses $P_A$ and $P_B$ describe the
individual detector excitations arising from their interactions
with the field, whereas $X$ quantifies the nonlocal coherence
generated between the two detector worldlines.

To leading order in perturbation theory, entanglement is harvested whenever the exchange correlations dominate the local detector noise. For identical detectors this condition reduces to

\begin{equation}
|X|
>
P,
\label{eq:HarvestCondition}
\end{equation}
which is equivalent to the positivity of the leading-order entanglement negativity. Equations~(\ref{eq:PGeneral}) and (\ref{eq:XGeneral}) constitute the central result of the present formalism. They establish a remarkable universality of the entanglement harvesting problem: the detector model, perturbative expansion, and entanglement criterion remain identical irrespective of the physical scenario. All dependence on detector trajectories, observer motion, spacetime geometry, and the quantum state of the field is encoded exclusively in the Wightman function appearing in the integral kernels. Once the appropriate two-point function has been specified, the evaluation of the detector response proceeds through exactly the same computational pipeline. This observation forms the foundation of the unified computational framework developed in the following section.

\subsection{Physical Interpretation and Computational Strategy}

The formalism developed above shows that entanglement harvesting is governed entirely by two quantities: the local excitation probability $P$ and the nonlocal exchange amplitude $X$. Although both originate from the same detector--field interaction, they describe fundamentally different physical processes.

The quantity $P$ characterizes the response of an individual detector to the quantum fluctuations of the field along its own worldline. Since it depends only on local field correlations, it represents detector excitation noise and remains nonzero even when the two detectors are infinitely separated. In contrast, the exchange amplitude $X$ arises from correlations between the field evaluated at two distinct spacetime points and therefore probes the nonlocal structure of the quantum field. It is this exchange term that enables two initially independent detectors to become entangled without any direct interaction between them.

The competition between these local and nonlocal processes
determines whether entanglement can be harvested. When the local
excitation probabilities of the two detectors are not necessarily
equal, the leading-order negativity of the reduced detector state
is given by

\begin{equation}
\mathcal{N}
=
\max\left\{
0,
\frac{
\sqrt{(P_A-P_B)^2+4|X|^2}
-(P_A+P_B)
}{2}
\right\},
\label{eq:Negativity}
\end{equation}

so that the corresponding harvesting condition is

\begin{equation}
|X|^2>P_AP_B.
\label{eq:HarvestCriterion}
\end{equation}

For identical local responses, $P_A=P_B\equiv P$, these
expressions reduce to the familiar result
$\mathcal{N}=\max(0,|X|-P)$, with harvesting occurring whenever
$|X|>P$. This condition provides a simple and physically
transparent interpretation of the harvesting process. The local
detector responses act as sources of noise, while the exchange
amplitude represents the nonlocal field correlations that generate
coherence between the detectors. Harvesting is therefore possible
only when the nonlocal correlations encoded in the quantum field
are sufficiently strong compared with the local excitation noise.

An equally important observation concerns the mathematical structure of the detector response. Equations~(\ref{eq:PGeneral}) and (\ref{eq:XGeneral}) show that the detector model, perturbative expansion, and entanglement criterion are completely independent of the underlying spacetime or observer motion. All information about the physical environment enters exclusively through the Wightman two-point function appearing in the integral kernels. Consequently, different physical situations correspond simply to different choices of the field correlation function,

\[
W(x,x')
\quad\longrightarrow\quad
\{P,X\}
\quad\longrightarrow\quad
\rho_{AB}
\quad\longrightarrow\quad
\mathcal{N}.
\]

This observation forms the conceptual foundation of the unified computational framework developed in the next section. Rather than constructing separate formalisms for inertial observers, uniformly accelerated observers, and thermal field states, we formulate a common computational pipeline in which the analytical reduction, numerical implementation, and entanglement diagnostics remain unchanged. The only ingredient that must be replaced from one physical setting to another is the corresponding Wightman function. This modular structure not only simplifies the calculations but also permits a direct comparison of harvesting behaviour across different observer trajectories and quantum field states within a single unified framework.

\section{A General Computational Framework}

The general formalism developed in the previous section expresses
entanglement harvesting entirely in terms of two detector response
functions: the local excitation probability $P$ and the nonlocal exchange
amplitude $X$. These quantities are determined by spacetime integrals over
the Wightman two-point function and are formally valid for arbitrary
detector trajectories and quantum states of the field.

Although the perturbative expressions are universal, their direct
evaluation is computationally demanding. The detector response is
described by highly oscillatory multidimensional integrals whose explicit
form depends on the underlying field correlations. Performing these
calculations independently for every spacetime or detector trajectory
would therefore obscure the common mathematical structure of the problem.

In this section we develop a unified computational framework that
separates the universal detector dynamics from the geometry-dependent
field correlations. The key observation is that the detector model,
switching function, perturbative expansion, and entanglement criterion
remain identical in every physical scenario considered in this work. The
only quantity that changes is the Wightman function describing the quantum
field. Once an efficient procedure has been established for evaluating the
detector response for a general two-point function, the same analytical
reduction and numerical implementation can be applied without
modification to inertial detectors, thermal field states, uniformly
accelerated observers, and more general backgrounds.

The present section therefore develops the computational machinery that
will be employed throughout the remainder of the paper. The subsequent
sections differ only through the specific form of the Wightman function,
while the detector dynamics and numerical pipeline remain unchanged.

\subsection{General Computational Strategy}

The detector response may be regarded as a functional of the Wightman
two-point function,

\begin{align}
P &= \mathcal{F}_{P}[W],\\
X &= \mathcal{F}_{X}[W],
\end{align}

where the functionals $\mathcal{F}_{P}$ and $\mathcal{F}_{X}$ are
completely determined by the detector model, switching profile, and the
leading-order perturbative expansion. This viewpoint provides a
particularly transparent interpretation of entanglement harvesting.
Different physical situations do not require new detector dynamics;
instead, they correspond simply to different realizations of the quantum
field correlations encoded in $W(x,x')$.

The computational problem therefore separates naturally into two stages.
The first consists of specifying the appropriate Wightman function for
the spacetime geometry, observer motion, and quantum state under
consideration. The second consists of evaluating the universal detector
functionals using a common analytical and numerical procedure. All
geometric information is therefore isolated within the two-point
correlation function.

The workflow adopted throughout this paper may be summarized
schematically as

\begin{equation}
\boxed{
\text{Spacetime}
\;\longrightarrow\;
W(x,x')
\;\longrightarrow\;
(P,X)
\;\longrightarrow\;
\rho_{AB}
\;\longrightarrow\;
\mathcal{N}.
}
\label{eq:workflow}
\end{equation}

Equation~(\ref{eq:workflow}) represents the central organizing principle
of the present work. Once the Wightman function has been specified, every
remaining stage of the calculation proceeds identically, irrespective of
the physical setting.

\subsection{Analytical Reduction of the Detector Response}

Although the perturbative expressions for $P$ and $X$ are exact to
leading order, their direct numerical evaluation is inefficient because
they involve oscillatory double integrals containing singular correlation
functions. A more efficient representation can be obtained by exploiting
the universal structure of the detector switching function before
introducing any spacetime-specific information.

We first introduce the relative and average proper-time coordinates

\begin{equation}
u=\tau-\tau',
\qquad
v=\frac{\tau+\tau'}{2},
\label{eq:uvGeneral}
\end{equation}

which separate the temporal separation of the interaction events from
their average interaction time. For Gaussian switching functions, the
switching profile factorizes exactly into independent functions of $u$
and $v$. Whenever the Wightman function depends only on the relative
spacetime separation, the integration over the average interaction time
can be carried out analytically, reducing the original two-dimensional
integrals to one-dimensional representations.

The remaining singular behaviour of the Wightman function is handled
through the distributional decomposition described in Appendix~A, where
the principal-value and local distributional contributions are separated
systematically. This procedure isolates the universal singular structure
common to all detector configurations while leaving the geometry-dependent
information entirely within the regular part of the correlation function.

The detector response therefore assumes the generic form

\begin{align}
P &= \lambda^{2}\int_{-\infty}^{\infty}du\,K_P(u),\\
X &= \lambda^{2}\int_{-\infty}^{\infty}du\,K_X(u),
\end{align}

where the kernels $K_P$ and $K_X$ contain the Gaussian switching profile,
oscillatory detector phases, and the appropriate Wightman function. Their
explicit forms depend on the physical scenario under consideration and
will be presented separately in the following sections.

Importantly, this reduction introduces no approximation beyond the
leading-order perturbative expansion itself. It merely reorganizes the
detector response into a mathematically equivalent form that is
considerably more efficient for numerical evaluation and provides a
common computational structure for every spacetime considered in this
work.

\subsection{Dimensionless Formulation and Numerical Implementation}

A further simplification is obtained by expressing the detector response
in terms of dimensionless variables. Besides improving numerical
stability, this formulation allows the harvesting behaviour to be
compared directly across different physical scales and spacetime
backgrounds.

The detector switching timescale $\sigma$ provides the natural unit of
time. For inertial detectors we therefore introduce

\begin{equation}
\alpha=\Omega\sigma,
\qquad
\beta=\frac{L}{\sigma},
\label{eq:DimensionlessMinkowski}
\end{equation}

where $\alpha$ characterizes the detector energy gap relative to the
interaction duration, while $\beta$ measures the detector separation in
units of the switching width.

Additional physical settings introduce only one extra dimensionless
parameter. Thermal field states are characterized by

\[
\theta=T\sigma,
\]

whereas uniformly accelerated detectors introduce

\[
\gamma=a\sigma.
\]

The detector response in every physical scenario is therefore described
by a small set of dimensionless control parameters.

The numerical procedure employed throughout this work is identical for
every spacetime considered:

\begin{enumerate}
\item Specify the appropriate Wightman function.
\item Construct the reduced one-dimensional detector kernels.
\item Evaluate the local excitation probability $P$ and exchange
amplitude $X$ numerically.
\item Construct the reduced detector density matrix and compute the
entanglement negativity.
\item Repeat the calculation over the relevant parameter space to obtain
harvesting landscapes and phase boundaries.
\end{enumerate}

The only stage that depends on the underlying spacetime is the choice of
the Wightman function. Every subsequent analytical reduction and
numerical calculation proceeds without modification. This modular
structure is the principal advantage of the present framework and enables
a direct comparison of entanglement harvesting across inequivalent vacuum
states, thermal fields, accelerated observers, and more general
background geometries. In the following sections we first apply this
framework to the Minkowski vacuum before extending the analysis to
thermal and accelerated detector configurations.

\section{Entanglement Harvesting in the Minkowski Vacuum}

We begin by applying the unified computational framework developed in the previous section to the simplest physical configuration: two inertial Unruh--DeWitt detectors interacting with a massless scalar field prepared in the Minkowski vacuum. Owing to its maximal spacetime symmetry and the absence of thermal or kinematic effects, this configuration provides the natural reference against which all subsequent results will be compared.

Besides serving as a benchmark for the analytical reduction and numerical implementation, the Minkowski vacuum establishes the baseline harvesting landscape arising solely from the interplay between local detector excitations and the intrinsic nonlocal correlations of the quantum vacuum. Reproducing this well-understood behaviour provides an important validation of the computational framework before extending the analysis to finite-temperature field states and uniformly accelerated observers.

In this section we first specify the detector trajectories and the corresponding Minkowski Wightman function. We then obtain reduced integral representations for the local excitation probability and exchange amplitude using the unified formalism developed in Sec.~III. Finally, we investigate the harvesting landscape over the relevant dimensionless parameter space, identifying the conditions under which vacuum entanglement can be extracted and establishing the benchmark results used throughout the remainder of this work.

\subsection{Physical Configuration}

We begin by considering the simplest detector configuration used throughout
this work, namely two identical Unruh--DeWitt detectors at rest in
four-dimensional Minkowski spacetime. This inertial vacuum configuration
serves as the benchmark against which the effects of finite temperature
and uniform acceleration will subsequently be compared.

The detectors are separated by a fixed distance $L$ along the $x$-axis
and follow the worldlines

\begin{align}
x_A^\mu(\tau)
&=
(\tau,-L/2,0,0),
\\
x_B^\mu(\tau)
&=
(\tau,+L/2,0,0),
\label{eq:MinkowskiWorldlines}
\end{align}

where the detector proper time coincides with the Minkowski coordinate
time. Both detectors possess the same energy gap $\Omega$ and interact
with the scalar field through the Gaussian switching function introduced
in Sec.~II.

The scalar field is prepared in the Minkowski vacuum,

\begin{equation}
|\Psi\rangle
=
|0_M\rangle,
\end{equation}

which is invariant under the Poincar\'e group and therefore represents
the vacuum perceived by inertial observers. The corresponding
positive-frequency Wightman function is

\begin{equation}
W_M(x,x')
=
-
\frac{1}{4\pi^2}
\frac{1}
{(t-t'-i\epsilon)^2-
|\mathbf{x}-\mathbf{x}'|^2},
\label{eq:MinkowskiWightman}
\end{equation}

where $\epsilon\rightarrow0^{+}$ implements the usual Feynman causal
prescription.

Since the detector trajectories are fixed, the detector response is
completely determined by evaluating the Wightman function along the
detector worldlines. For interactions occurring on a single detector
trajectory, the spatial separation vanishes, yielding the local
correlation function

\begin{equation}
W_M^{\mathrm{loc}}(\tau,\tau')
=
-
\frac{1}{4\pi^2}
\frac{1}
{(\tau-\tau'-i\epsilon)^2},
\label{eq:WMlocal}
\end{equation}

whereas the field correlations connecting the two detector worldlines are

\begin{equation}
W_M^{\mathrm{nonloc}}(\tau,\tau')
=
-
\frac{1}{4\pi^2}
\frac{1}
{(\tau-\tau'-i\epsilon)^2-L^2}.
\label{eq:WMnonlocal}
\end{equation}

These local and nonlocal correlation functions constitute the only
physical input required for evaluating the detector excitation
probability and the exchange amplitude introduced in Sec.~II. The former
describes the unavoidable local vacuum fluctuations sampled by each
detector, while the latter encodes the nonlocal field correlations
responsible for entanglement harvesting.

In the following subsection we apply the unified computational framework
developed in Sec.~III to these Minkowski correlation functions. The
resulting analytical reduction provides numerically efficient
one-dimensional integral representations that form the basis of all
subsequent calculations presented in this work.

\subsection{Reduction of the Detector Response}

Having specified the detector trajectories and the corresponding
Minkowski Wightman function, we now evaluate the detector response
within the unified computational framework developed in Sec.~III.
Substituting Eqs.~(\ref{eq:WMlocal}) and
(\ref{eq:WMnonlocal}) into the general expressions,
Eqs.~(\ref{eq:PGeneral}) and (\ref{eq:XGeneral}),
yields the local excitation probability and exchange amplitude for two
inertial detectors interacting with the Minkowski vacuum.

A direct evaluation of these expressions involves oscillatory double
integrals over the detector proper times. As shown in Sec.~III,
however, the detector response possesses a universal structure that
allows these integrals to be reduced analytically before numerical
evaluation. Introducing average and relative proper-time coordinates,
the Gaussian switching functions factorize, permitting the integration
over the average interaction time to be performed exactly. The remaining
expressions depend only on the relative proper time and are therefore
reduced to one-dimensional integral representations. The complete
derivation, including the treatment of the distributional singularities
of the Wightman function, is presented in Appendix~A.

The resulting local detector response may be written as

\begin{equation}
P
=
\mathcal{P}(\Omega,\sigma),
\label{eq:PMinkowskiReduced}
\end{equation}

while the exchange amplitude assumes the form

\begin{equation}
X
=
\mathcal{X}(\Omega,L,\sigma),
\label{eq:XMinkowskiReduced}
\end{equation}

where the explicit one-dimensional integral expressions are given in
Appendix~A. These reduced representations are mathematically equivalent
to the original double-integral formulation but are considerably more
efficient for numerical computation.

An important feature of this reduction is that no additional
approximation is introduced beyond the leading-order perturbative
expansion. The transformation from double to one-dimensional integrals
is exact and preserves the complete detector dynamics encoded in the
Minkowski Wightman function. Consequently, the detector response
obtained here provides the benchmark implementation of the computational
framework developed in Sec.~III.

Moreover, the reduction itself is independent of the particular quantum
state of the field. Once the appropriate Wightman function is specified,
the analytical reduction and subsequent numerical evaluation proceed in
exactly the same manner. This modular structure enables the detector
responses for the Minkowski vacuum, thermal field states, and uniformly
accelerated observers to be computed within a common framework, allowing
their harvesting behaviour to be compared directly.

In the following subsection, these reduced detector response functions
are evaluated over the dimensionless parameter space introduced in
Sec.~III to construct the benchmark harvesting landscape for the
Minkowski vacuum.

\subsection{Harvesting Landscape and Benchmark Behaviour}

We now employ the reduced detector response functions derived in the previous subsection to investigate entanglement harvesting in the Minkowski vacuum. Throughout this analysis, the detector configuration is completely specified by the dimensionless parameters introduced in Sec.~III,

\[
\alpha=\Omega\sigma,
\qquad
\beta=\frac{L}{\sigma},
\]

which characterize the detector energy gap relative to the interaction duration and the detector separation relative to the switching width, respectively. The harvesting behaviour is therefore described entirely within the two-dimensional parameter space $(\alpha,\beta)$.

For each point in this parameter space, the one-dimensional
integral representations are evaluated numerically to obtain
the local detector excitation probabilities $P_A$ and $P_B$,
together with the exchange amplitude $X$. These quantities
determine the reduced detector density matrix through
Eq.~(\ref{eq:DensityMatrix}), from which the entanglement
negativity is computed using the general expression established
in Sec.~II. Since this represents the simplest physical setting
considered in the present work, the resulting harvesting
landscape serves as the benchmark against which the effects
of finite temperature and uniform acceleration will
subsequently be compared.

Entanglement harvesting arises from the competition between two
physically distinct processes. The local detector response reflects
the vacuum fluctuations sampled by each detector, while the exchange
amplitude measures the nonlocal quantum correlations of the field
shared between the two detector worldlines. For the present
configuration of identical inertial detectors, harvesting occurs
only when these nonlocal correlations dominate the local detector
noise,

\begin{equation}
|X|>P,
\label{eq:HarvestConditionMinkowski}
\end{equation}
which defines the boundary separating harvesting and
non-harvesting regions throughout the parameter space.

\begin{figure}
    \centering
    \includegraphics[width=0.5\linewidth]{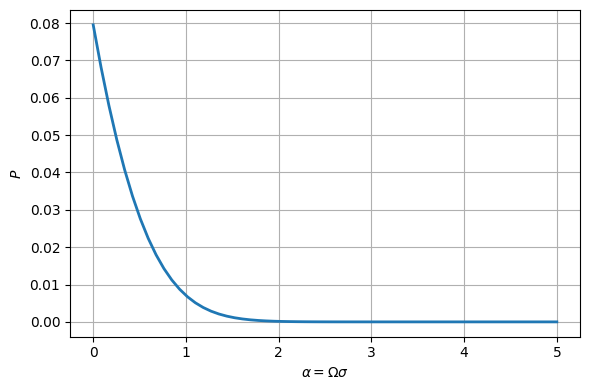}
    \caption{
Local detector excitation probability $P(\alpha)$ in the
Minkowski vacuum as a function of the dimensionless detector
energy gap $\alpha=\Omega\sigma$, for a fixed coupling
$\lambda=1$. The response decreases monotonically with
increasing energy gap and becomes strongly suppressed in the
large-gap regime.
}
    \label{fig:MinkowskiP}
\end{figure}

Figure~\ref{fig:MinkowskiP} shows the local detector excitation
probability as a function of the dimensionless detector gap
$\alpha=\Omega\sigma$. The excitation probability decreases
rapidly with increasing $\alpha$, reflecting the reduced ability
of large-gap detectors to absorb vacuum fluctuations during the
finite interaction time. Since the local response depends only
on the individual detector trajectory, it is independent of
the detector separation. Furthermore, according to Eq.~(A12),
the response depends on the dimensionless gap $\alpha$ and the
coupling $\lambda$, but not independently on the switching width
$\sigma$. Consequently, at fixed $\alpha$ and $\lambda$, the
excitation probability is identical for different values of
$\sigma$.

\begin{figure}
    \centering
    \includegraphics[width=0.5\linewidth]{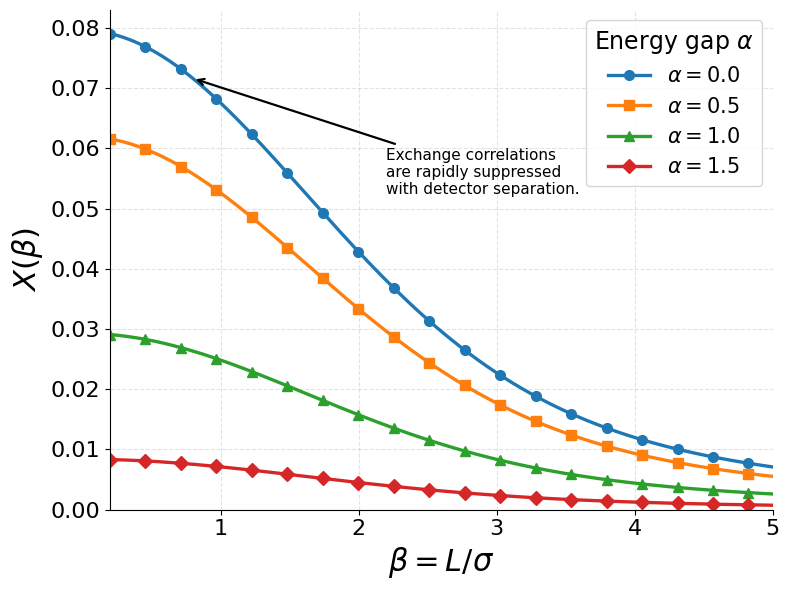}
    \caption{
Exchange amplitude $|X|$ as a function of the dimensionless detector separation $\beta=L/\sigma$ for representative detector energy gaps $\alpha=\Omega\sigma$ in the Minkowski vacuum. The exchange correlations decrease monotonically with increasing detector separation and are progressively suppressed for larger detector energy gaps, reflecting the rapid decay of nonlocal vacuum correlations responsible for entanglement harvesting.
}
    \label{fig:MinkowskiX}
\end{figure}

The corresponding exchange amplitude is shown in Fig.~\ref{fig:MinkowskiX}. Unlike the local detector response, the exchange amplitude depends
explicitly on the spatial separation because it measures the
nonlocal vacuum correlations shared by the two detector worldlines.
As these correlations decay with distance, the exchange amplitude
is rapidly suppressed beyond separations comparable to the switching
width. Larger detector gaps further reduce the exchange process,
since the rapidly oscillating detector phase increasingly averages
out the field correlations during the interaction.

\begin{figure}
    \centering
    \includegraphics[width=0.5\linewidth]{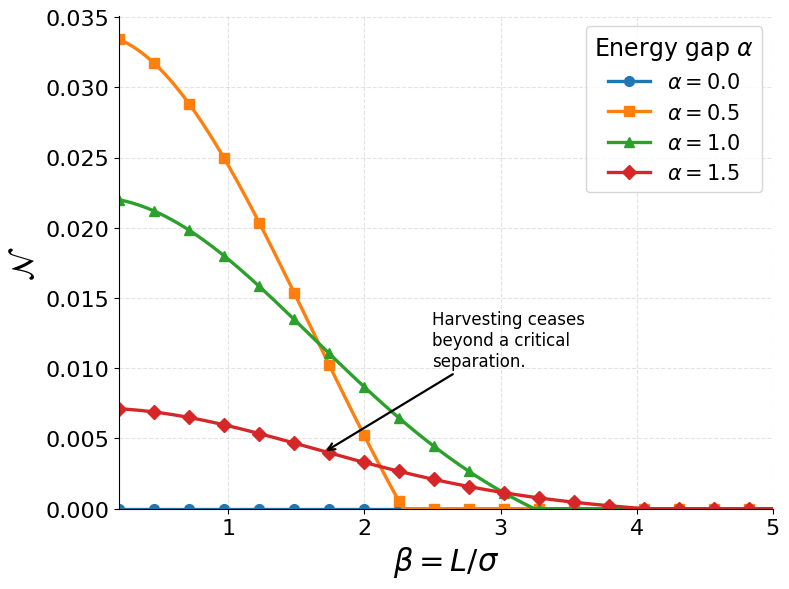}
    \caption{
    Harvested entanglement negativity $\mathcal{N}$ as a function
    of the dimensionless detector separation $\beta=L/\sigma$
    for representative detector energy gaps $\alpha=\Omega\sigma$
    in the Minkowski vacuum. For each energy gap, the harvested
    entanglement decreases with increasing separation and vanishes
    beyond a critical distance, reflecting the weakening of
    nonlocal vacuum correlations. The existence of an optimal
    intermediate detector gap ($\alpha\approx0.5$) demonstrates
    that efficient entanglement harvesting results from a balance
    between local detector noise and nonlocal field correlations.
    }
    \label{fig:MinkowskiN}
\end{figure}

Combining the local excitation probability and exchange amplitude
yields the harvested entanglement negativity shown in
Fig.~\ref{fig:MinkowskiN}. The negativity exhibits a well-defined
maximum at an intermediate detector gap for the parameter range
considered, indicating that harvesting is optimized when the
balance between local detector noise and nonlocal vacuum
correlations is most favourable. For sufficiently large detector
gaps, the harvested entanglement decreases rapidly as the
nonlocal exchange amplitude becomes increasingly suppressed.

\begin{figure}
    \centering
    \includegraphics[width=0.5\linewidth]{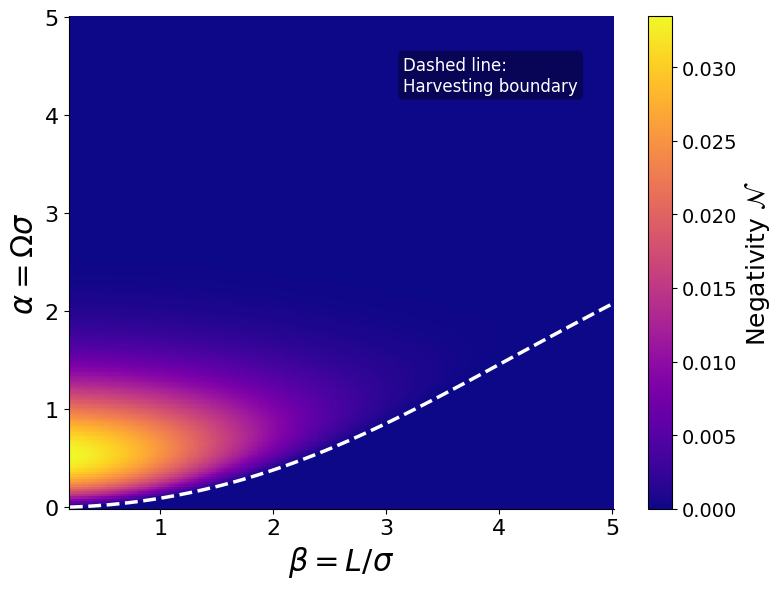}
    \caption{
    Harvesting landscape in the Minkowski vacuum showing the
    entanglement negativity $\mathcal{N}$ over the parameter space
    $(\alpha,\beta)$, where $\alpha=\Omega\sigma$ and
    $\beta=L/\sigma$. The colour scale represents the magnitude
    of the harvested entanglement, with brighter regions
    corresponding to stronger harvesting. The dashed curve denotes
    the harvesting boundary ($\mathcal{N}=0$, equivalently
    $|X|=P$), separating the entangled and separable regions.
    Maximum harvesting occurs for small detector separations and
    intermediate detector gaps, while increasing the detector gap
    or separation eventually suppresses the nonlocal correlations
    required for entanglement extraction.
    }
    \label{fig:MinkowskiLandscape}
\end{figure}

The complete harvesting landscape is presented in
Fig.~\ref{fig:MinkowskiLandscape}. The colour scale represents
the harvested entanglement negativity over the full
$(\alpha,\beta)$ parameter space, while the white contour denotes
the harvesting threshold defined by $|X|=P$. Harvesting is
confined to a finite region of sufficiently small detector
separations and intermediate detector gaps, beyond which the
nonlocal exchange correlations are unable to overcome the local
detector excitation probability. The landscape clearly illustrates
that efficient harvesting requires both sufficiently strong vacuum
correlations and an appropriate detector energy gap. While reducing
the detector separation generally enhances the exchange
correlation, small detector gaps also produce a relatively large
local excitation probability, preventing the harvested entanglement
from growing indefinitely. The brightest region of the landscape
therefore occurs at an intermediate detector gap rather than at
$\alpha=0$, demonstrating that optimal harvesting results from
a balance between local excitation and nonlocal correlation.

\begin{figure}
    \centering
    \includegraphics[width=0.5\linewidth]{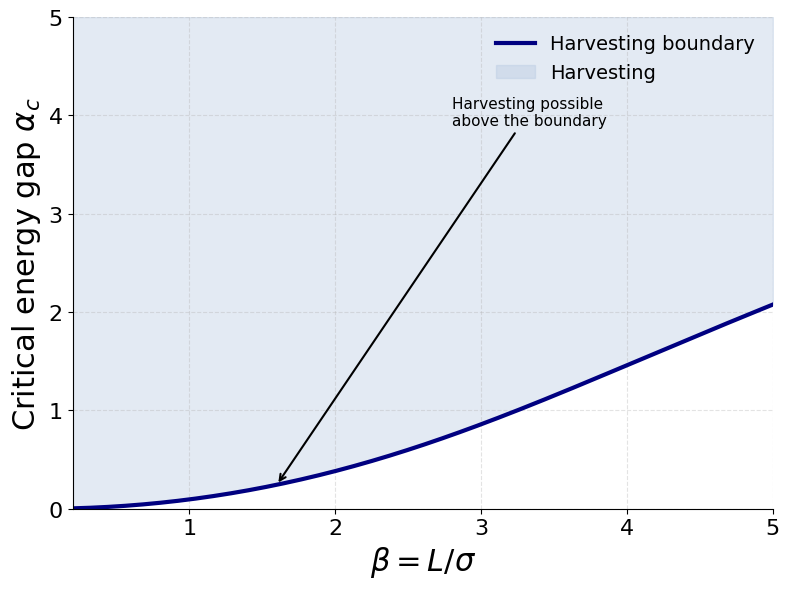}
    \caption{
    Critical harvesting boundary in the Minkowski vacuum.
    The solid curve represents the critical detector gap
    $\alpha_c=\Omega_c\sigma$ obtained from the condition
    $|X|=P$, separating the harvesting and non-harvesting
    regions of the $(\beta,\alpha)$ parameter space.
    Detector configurations satisfying $|X|>P$ permit
    entanglement harvesting, whereas those satisfying
    $|X|\leq P$ do not. This boundary provides the benchmark
    against which the corresponding finite-temperature and
    uniformly accelerated harvesting regions are compared
    in the following sections.
    }
    \label{fig:MinkowskiBoundary}
\end{figure}

The harvesting boundary is extracted separately in
Fig.~\ref{fig:MinkowskiBoundary}, providing a concise
representation of the transition between harvesting and
non-harvesting regions. This boundary defines the critical
detector separation associated with the harvesting condition
as a function of the detector gap and serves as the principal
benchmark for the corresponding boundaries obtained in
finite-temperature and uniformly accelerated field
configurations. The boundary summarizes how the competition
between local detector excitation and nonlocal vacuum
correlations determines the range of detector separations
over which entanglement can be extracted. It therefore
provides a compact representation of the complete harvesting
landscape and will prove particularly useful when comparing
how thermal fluctuations and uniform acceleration deform
the harvesting region.

The Minkowski vacuum therefore establishes the reference
harvesting landscape against which the effects of finite
temperature and uniform acceleration may be assessed. In
the absence of thermal or kinematic modifications, the
competition between local detector noise and nonlocal vacuum
correlations produces a well-defined harvesting region
bounded by Eq.~(\ref{eq:HarvestConditionMinkowski}). The
following sections employ exactly the same computational
framework to investigate how this benchmark landscape is
modified when the field correlations are altered by finite
temperature and observer acceleration. This direct comparison
enables the effects of thermal fluctuations and observer motion
to be isolated without changing the underlying detector model
or perturbative formalism.

\section{Entanglement Harvesting in Thermal Field States}

Having established the harvesting behaviour for two inertial detectors interacting with the Minkowski vacuum, we now investigate how the extraction of entanglement is modified when the scalar field is prepared in a thermal equilibrium state. In contrast to the vacuum, a thermal field contains real excitations that alter both the local detector response and the nonlocal field correlations responsible for entanglement harvesting. Understanding how these competing effects reshape the harvesting landscape is important not only for harvesting in finite-temperature environments but also because thermal field states provide the natural benchmark for comparison with uniformly accelerated detectors through the Unruh effect.

One of the principal advantages of the unified computational framework developed in Sec.~III is that the extension to thermal field states requires only a single modification: the Minkowski vacuum Wightman function is replaced by its finite-temperature counterpart. The detector model, Gaussian switching functions, perturbative expansion, analytical reduction, and numerical implementation remain unchanged. Consequently, any differences in the detector response, exchange amplitude, or harvested entanglement originate exclusively from the altered structure of the thermal field correlations, allowing the influence of temperature to be isolated unambiguously.

In this section we first derive the thermal field correlation functions relevant for inertial Unruh--DeWitt detectors before obtaining the corresponding detector response and exchange amplitude using the same reduction procedure employed for the Minkowski vacuum. The detailed analytical derivation is presented in Appendix~B. We then investigate how increasing temperature modifies the local detector noise, the nonlocal exchange correlations, and ultimately the harvesting landscape, thereby establishing the thermal reference against which the accelerated detector results presented in the following section will be compared.


\subsection{Thermal Field Correlations}

We consider the same pair of inertial Unruh--DeWitt detectors introduced in the previous section. Their worldlines, energy gap, spatial separation, and Gaussian switching function therefore remain unchanged,

\begin{align}
x_A^\mu(\tau)
&=
(\tau,-L/2,0,0),
\\
x_B^\mu(\tau)
&=
(\tau,+L/2,0,0),
\end{align}

allowing the effects of finite temperature to be isolated from those associated with detector motion or spacetime geometry.

The scalar field is assumed to be in a thermal equilibrium state described by the density operator

\begin{equation}
\rho_{\mathrm{th}}
=
\frac{e^{-\beta_T H}}
{\mathrm{Tr}\!\left(e^{-\beta_T H}\right)},
\label{eq:ThermalDensityMatrix}
\end{equation}

where $H$ denotes the free-field Hamiltonian,

\begin{equation}
\beta_T=\frac{1}{k_B T},
\end{equation}

is the inverse temperature. Throughout this work we employ natural units
$(\hbar=c=k_B=1)$ so that $\beta_T=1/T$.

The corresponding thermal Wightman function is defined by

\begin{equation}
W_T(x,x')
=
\mathrm{Tr}
\!\left[
\rho_{\mathrm{th}}
\,
\phi(x)
\phi(x')
\right].
\label{eq:ThermalWightmanGeneral}
\end{equation}

For thermal equilibrium states the correlation function satisfies the Kubo--Martin--Schwinger (KMS) condition,

\begin{equation}
W_T(\Delta t)
=
W_T(\Delta t+i\beta_T),
\end{equation}

which characterizes thermal quantum field theory and replaces the Lorentz-invariant vacuum correlations of the Minkowski state.

A convenient representation of the finite-temperature Wightman function is obtained by summing the zero-temperature correlator over its imaginary-time images,

\begin{equation}
W_T(x,x')
=
\sum_{n=-\infty}^{\infty}
W_M
\!\left(
t-t'-in\beta_T,
\mathbf{x}-\mathbf{x}'
\right),
\label{eq:MatsubaraRepresentation}
\end{equation}

which automatically satisfies the KMS condition. The explicit evaluation of this series is presented in Appendix~B.

For coincident spatial points one obtains

\begin{equation}
W_T^{\mathrm{loc}}(\Delta t)
=
-
\frac{T^2}{4}
\,
\mathrm{csch}^{2}
\!\left[
\pi T
(\Delta t-i\epsilon)
\right],
\label{eq:ThermalWightman}
\end{equation}

while for two detectors separated by a fixed distance $L$ the corresponding nonlocal correlation function becomes

\begin{equation}
W_T^{\mathrm{nonloc}}(\Delta t,L)
=
-
\frac{T}{8\pi L}
\left[
\coth\!\left(
\pi T(\Delta t-L-i\epsilon)
\right)
-
\coth\!\left(
\pi T(\Delta t+L-i\epsilon)
\right)
\right].
\label{eq:ThermalSeparated}
\end{equation}

The coincidence limit of Eq.~(\ref{eq:ThermalSeparated}) reproduces Eq.~(\ref{eq:ThermalWightman}), providing an important consistency check on the thermal correlation function.

Substituting Eqs.~(\ref{eq:ThermalWightman}) and (\ref{eq:ThermalSeparated}) into the general expressions derived in Sec.~II yields the excitation probability and exchange amplitude for inertial detectors immersed in a thermal field. Since the detector trajectories are identical to those considered in the Minkowski vacuum, every modification to the harvesting behaviour originates exclusively from the altered structure of the thermal field correlations. The resulting one-dimensional integral representations, derived in Appendix~B, provide the starting point for the numerical analysis presented in the following subsection.

\subsection{Reduction of the Detector Response}

The detector response for a thermal field state is obtained by evaluating the general expressions for the local excitation probability and exchange amplitude, Eqs.~(\ref{eq:PGeneral}) and (\ref{eq:XGeneral}), using the finite-temperature Wightman function derived in the previous subsection. Since the detector trajectories, Gaussian switching functions, and perturbative interaction remain identical to those considered in the Minkowski vacuum, the analytical reduction developed in Sec.~III carries over without modification.

Introducing the average and relative detector proper times and performing the Gaussian integration over the average interaction time analytically reduces both detector observables to one-dimensional integral representations. The only modification relative to the vacuum calculation is the replacement of the Minkowski Wightman function by its thermal counterpart. Consequently, the reduced detector response is completely determined by the thermal field correlations, while the overall computational framework remains unchanged.

The resulting local detector response may therefore be written as

\begin{equation}
P_T
=
\mathcal{P}_T(\Omega,L,\sigma,T),
\label{eq:PThermal}
\end{equation}

and the corresponding exchange amplitude as

\begin{equation}
X_T
=
\mathcal{X}_T(\Omega,L,\sigma,T),
\label{eq:XThermal}
\end{equation}

where the explicit one-dimensional integral expressions are obtained directly from the reduction procedure presented in Appendix~B after replacing the vacuum Wightman function by the thermal correlation function.

An important advantage of this formulation is that it isolates the physical effects of finite temperature in a transparent manner. The detector trajectories, switching functions, perturbative expansion, and numerical implementation are identical to those employed in the Minkowski benchmark. Consequently, any differences in the detector response and the resulting harvesting behaviour originate solely from the modification of the underlying field correlations by the thermal Bose--Einstein distribution.

The reduced detector response obtained here forms the basis for the numerical analysis presented in the following subsection, where the evolution of the harvesting landscape is investigated systematically as a function of the field temperature.

\subsection{Harvesting Landscape at Finite Temperature}

We now employ the reduced detector response functions derived in the previous subsection to investigate entanglement harvesting in thermal field states. As in the Minkowski vacuum analysis, the detector configuration is characterized by the dimensionless parameters

\[
\alpha=\Omega\sigma,
\qquad
\beta=\frac{L}{\sigma},
\]

together with the dimensionless temperature

\begin{equation}
\theta=T\sigma,
\label{eq:Theta}
\end{equation}

which measures the thermal energy scale relative to the detector interaction time. The harvesting problem is therefore completely specified by the three-dimensional parameter space $(\alpha,\beta,\theta)$.

For each temperature, the one-dimensional integral expressions derived in Appendix~B are evaluated numerically to obtain the local detector response $P_T$, the exchange amplitude $X_T$, and the corresponding entanglement negativity. Repeating this procedure for different temperatures allows the complete harvesting landscape to be followed continuously from the Minkowski vacuum to finite-temperature field states.

As in the vacuum case, harvesting occurs whenever the nonlocal exchange correlation exceeds the local detector response,

\begin{equation}
|X_T|>P_T,
\label{eq:ThermalHarvestCondition}
\end{equation}

which defines the boundary separating the harvesting and non-harvesting regions. Unlike the vacuum state, however, finite temperature modifies both the local detector noise and the nonlocal field correlations simultaneously. The numerical results presented below therefore determine whether thermal fluctuations suppress harvesting through increased detector noise or enhance harvesting through stronger nonlocal field correlations.

\begin{figure}
    \centering
    \includegraphics[width=0.5\linewidth]{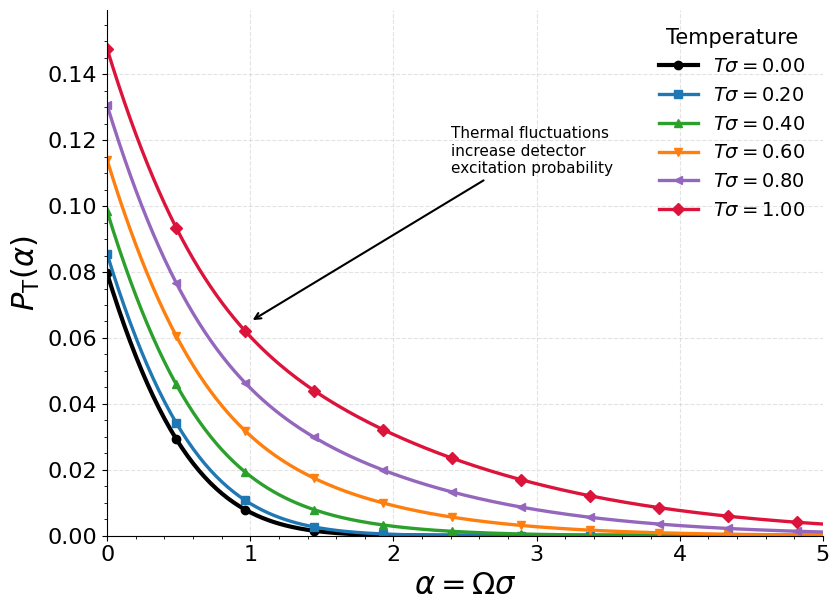}
    \caption{
    Thermal detector response $P_T(\alpha)$ as a function of the dimensionless energy gap $\alpha=\Omega\sigma$, for different values of the dimensionless temperature $T\sigma$. The vacuum response ($T\sigma=0$) is recovered smoothly in the zero-temperature limit. Increasing temperature enhances the detector excitation probability, resulting in a larger local response over the parameter range considered.
    }
    \label{fig:ThermalP}
\end{figure}

Figure~\ref{fig:ThermalP} shows the local detector response as a function of the detector energy gap for representative temperatures.

As expected, the detector excitation probability increases monotonically with temperature owing to the increased thermal occupation of field modes. The zero-temperature curve reproduces the Minkowski vacuum result obtained in the previous section. Thus, thermal fluctuations increase the local detector noise that competes with entanglement harvesting.

\begin{figure}
    \centering
    \includegraphics[width=0.5\linewidth]{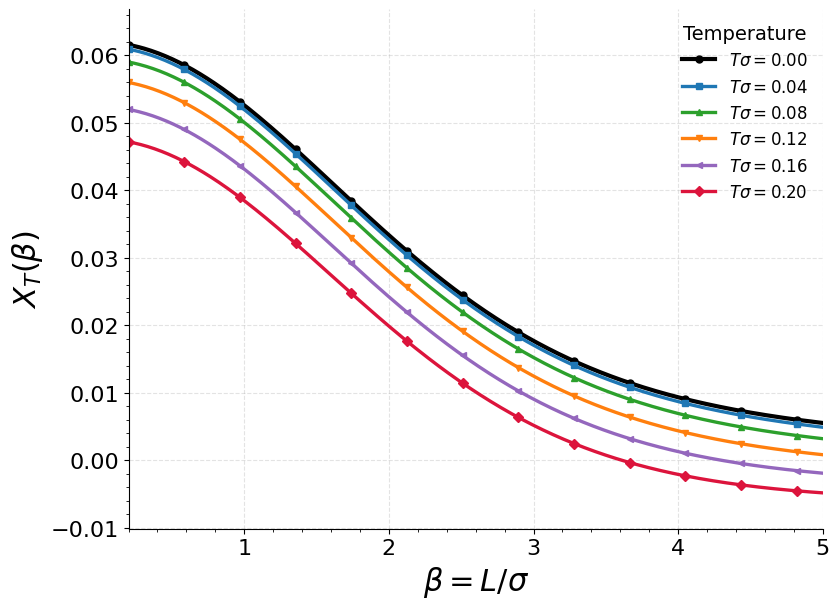}
    \caption{
    Thermal exchange amplitude $X_T(\beta)$ as a function of the dimensionless detector separation $\beta=L/\sigma$, for a fixed detector energy gap $\alpha=\Omega\sigma=0.5$ and different values of the dimensionless temperature $T\sigma$. The exchange amplitude decreases with increasing detector separation, while increasing temperature generally suppresses its magnitude and can drive it through zero at sufficiently large separations.
    }
    \label{fig:ThermalX}
\end{figure}

Figure~\ref{fig:ThermalX} presents the corresponding exchange amplitude.

The exchange amplitude generally decreases with detector separation and is suppressed as the temperature increases. Within the present Gaussian-switching perturbative framework, this indicates that thermal fluctuations weaken the nonlocal correlations contributing to entanglement harvesting. Whether harvesting persists, however, depends on the competition between this suppression and the temperature-dependent local detector response.

\begin{figure}
    \centering
    \includegraphics[width=0.5\linewidth]{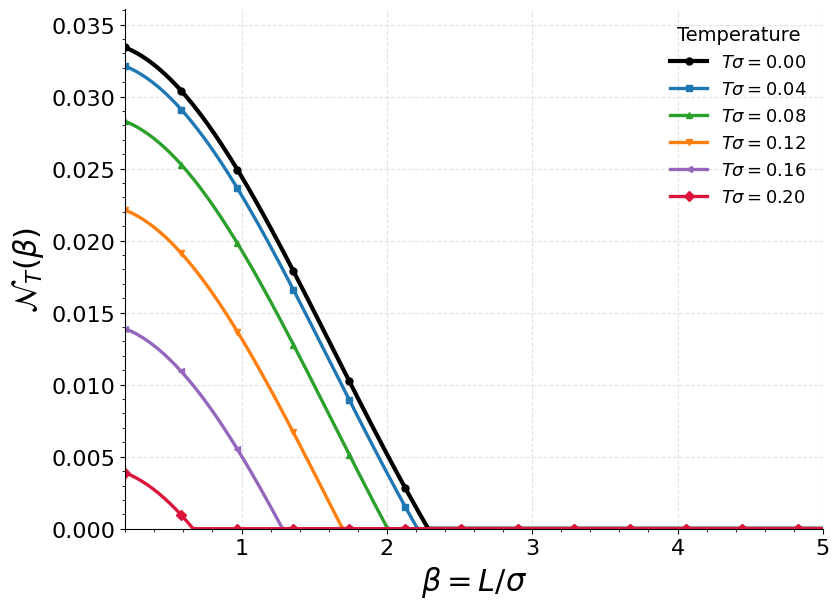}
    \caption{
    Harvested entanglement negativity $\mathcal{N}_T(\beta)$ as a function of the dimensionless detector separation $\beta=L/\sigma$, for a fixed detector energy gap $\alpha=\Omega\sigma=0.5$ and different values of the dimensionless temperature $T\sigma$. The negativity decreases with increasing detector separation and vanishes beyond a temperature-dependent critical separation. Increasing temperature suppresses the harvested entanglement and reduces the range over which entanglement harvesting occurs; for sufficiently high temperatures, the negativity vanishes throughout the plotted range.
    }
    \label{fig:ThermalNegativityBeta}
\end{figure}

The combined effect of these competing contributions is illustrated directly in Fig.~\ref{fig:ThermalNegativityBeta}.
Throughout the parameter range considered, the suppression of the exchange amplitude, together with the increase in local detector excitations, leads to a suppression of harvested entanglement. Consequently, the range of detector separations over which the negativity remains nonzero becomes progressively smaller. This demonstrates that thermal fluctuations reduce both the magnitude and the spatial range of entanglement harvesting.

\begin{figure}
    \centering
    \includegraphics[width=0.5\linewidth]{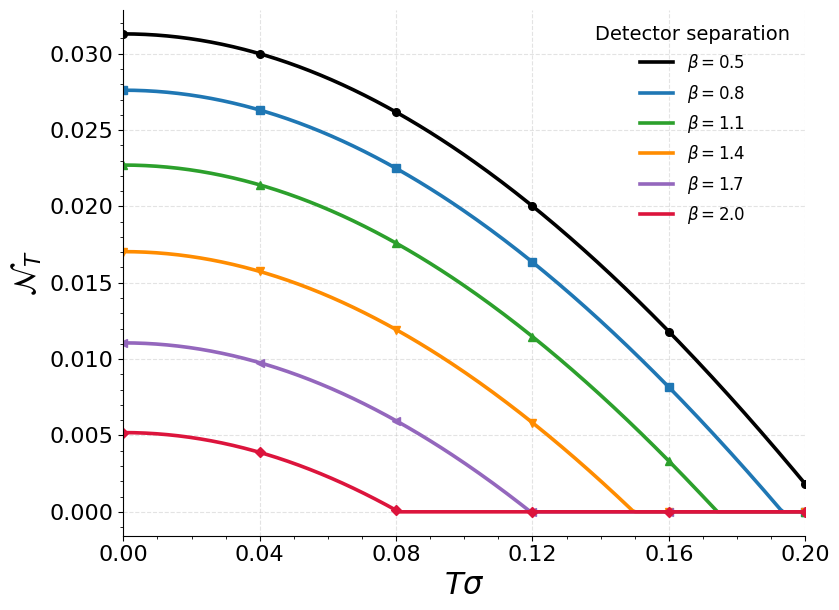}
    \caption{
    Harvested entanglement negativity $\mathcal{N}_T$ as a function of the dimensionless temperature $T\sigma$, for a fixed detector energy gap $\alpha=\Omega\sigma=0.5$ and different detector separations $\beta=L/\sigma$. The negativity decreases with increasing temperature and vanishes above a separation-dependent critical temperature. As the detector separation increases, the initial harvested entanglement decreases and entanglement harvesting ceases at progressively lower temperatures.
    }
    \label{fig:ThermalNegativityTemperature}
\end{figure}

The temperature dependence of the harvested entanglement is examined further in Fig.~\ref{fig:ThermalNegativityTemperature}.

For every detector separation considered, the negativity decreases monotonically with temperature, with the suppression becoming increasingly pronounced at larger separations. This indicates that thermal effects are particularly detrimental to entanglement harvesting at larger detector separations, where the nonlocal correlations are already weaker.

\begin{figure}
    \centering
    \includegraphics[width=0.5\linewidth]{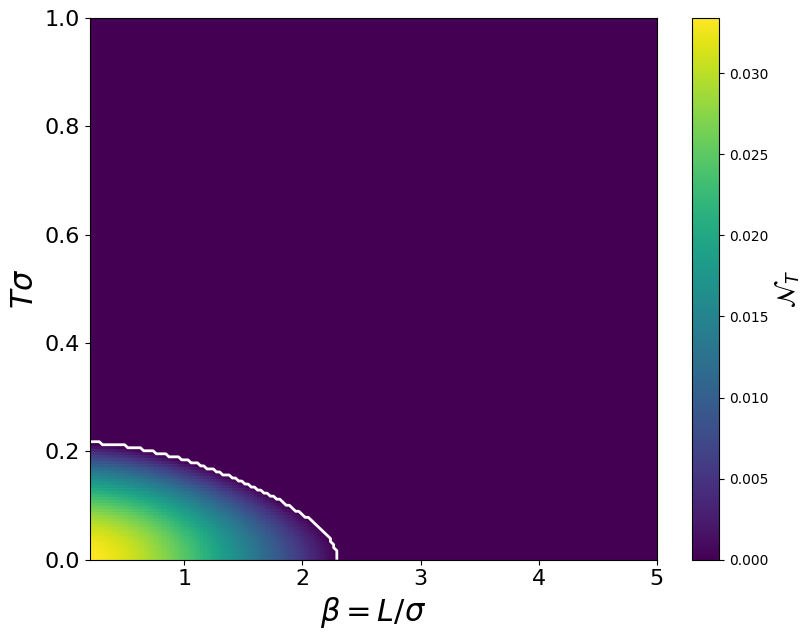}
    \caption{
    Phase diagram of harvested entanglement negativity $\mathcal{N}_T$ as a function of the dimensionless detector separation $\beta=L/\sigma$ and temperature $T\sigma$, for a fixed detector energy gap $\alpha=\Omega\sigma=0.5$. The colour scale represents the magnitude of the negativity, while the white contour marks the boundary $\mathcal{N}_T=0$ separating the entanglement-harvesting and non-harvesting regions. Harvested entanglement is suppressed by increasing temperature and detector separation, surviving only within a restricted region of parameter space.
    }
    \label{fig:ThermalLandscape}
\end{figure}

The harvesting landscape at fixed detector energy gap is summarized in Fig.~\ref{fig:ThermalLandscape}.

The phase diagram provides a global visualization of the harvesting process at the chosen detector energy gap. The colour scale represents the magnitude of the harvested entanglement, while the white contour marks the harvesting boundary separating the entangled and separable regions. As the temperature increases, the harvesting boundary shifts toward progressively smaller detector separations, demonstrating that finite temperature reduces the region of parameter space over which harvesting remains possible. Thus, within the perturbative regime considered here, thermal fluctuations suppress both the magnitude of the harvested entanglement and the range of separations over which it can be extracted.

Finally, we investigated whether the detector energy gap exhibits an analogous critical behaviour. Specifically, we searched for a temperature-dependent critical gap $\alpha_c=\Omega_c\sigma$ defined by the condition
$\mathcal{N}_T(\alpha_c,\beta,\theta)=0$. Within the parameter range explored in the present perturbative analysis, no finite critical value was observed. Instead, the harvested entanglement decreases smoothly with increasing detector energy gap, without exhibiting a sharp transition in the range considered. The principal effect of finite temperature is therefore to modify the balance between local detector excitations and nonlocal field correlations, while leaving the qualitative dependence on the detector energy gap unchanged over the parameter range explored.

\section{Entanglement Harvesting for Uniformly Accelerated Detectors}

We now turn to entanglement harvesting by uniformly accelerated
observers. Unlike the previous sections, where the detectors followed
inertial trajectories through either the Minkowski vacuum or a thermal
field state, the quantum field here remains in the Minkowski vacuum,
while the detector trajectories become noninertial. Consequently, the
field correlations sampled by the detectors are modified entirely by
their motion, providing a direct means of investigating how acceleration
influences the extraction of vacuum entanglement.

Uniformly accelerated detectors play a fundamental role in relativistic
quantum field theory through the Unruh effect, according to which a
uniformly accelerated observer perceives the Minkowski vacuum as a
thermal state. Entanglement harvesting offers a particularly sensitive
probe of this phenomenon, since it depends not only on the local
excitation probability of each detector but also on the nonlocal field
correlations responsible for the exchange of quantum information between
the detectors. It therefore provides an ideal setting in which to
examine the similarities and differences between acceleration-induced
and thermal effects.

Within the unified computational framework developed in Sec.~III, the
transition from inertial to accelerated detectors again requires only a
single modification: the detector trajectories and the corresponding
Wightman function. The perturbative formalism and numerical
implementation remain unchanged. This enables a direct comparison
between harvesting in the Minkowski vacuum, finite-temperature field
states, and uniformly accelerated detector configurations, allowing the
influence of observer motion to be isolated from that of genuine thermal
fluctuations.

In this section we first introduce the trajectories of uniformly
accelerated detectors and the corresponding field correlation function.
We then evaluate the detector response within the same computational
framework employed throughout this work before investigating the
resulting harvesting landscape and comparing it directly with the
Minkowski vacuum and thermal cases.

\subsection{Uniformly Accelerated Detector Trajectories}

We consider two identical Unruh--DeWitt detectors undergoing Born-rigid
uniform acceleration in the $x$-direction. Although the detectors are
identical in their internal properties, their proper accelerations differ
because they occupy different positions in the accelerated frame.
Throughout this section, the quantum field remains in the Minkowski
vacuum,

\begin{equation}
|\Psi\rangle = |0_M\rangle,
\end{equation}

so that all modifications to the detector response arise from the
noninertial motion of the detectors.

For definiteness, the detectors accelerate uniformly along the
$x$-direction while maintaining a constant proper separation. We denote
the proper acceleration of detector $A$ by $a$ and their constant
Rindler separation by $L_R$. Their proper accelerations are
\begin{equation}
a_A=a,
\qquad
a_B=\frac{a}{1+aL_R}.
\end{equation}
The corresponding proper times are also different. If $\eta$ denotes
the common Rindler time, then
\begin{equation}
\tau_A=\eta,
\qquad
\tau_B=(1+aL_R)\eta.
\end{equation}
The explicit trajectories and their proper-time parametrizations are
derived in Appendix~C.

The positive-frequency Wightman function is still that of the
Minkowski vacuum,

\begin{equation}
W_M(x,x')
=
-\frac{1}{4\pi^2}
\frac{1}{
(t-t'-i\epsilon)^2-
|\mathbf{x}-\mathbf{x}'|^2
},
\label{eq:AcceleratedWightman}
\end{equation}

but its evaluation along the accelerated trajectories produces
correlation functions that differ from those sampled by inertial
detectors. In particular, the invariant spacetime interval becomes a
nonlinear function of the detector proper times, reflecting the
hyperbolic geometry of the accelerated motion and leading to modified
local and nonlocal detector correlations.

Substituting the Born-rigid trajectories into
Eq.~(\ref{eq:AcceleratedWightman}) yields the accelerated Wightman
functions governing the excitation probabilities and exchange amplitude.
Their explicit derivation is presented in Appendix~C. The resulting
expressions enter the detector response through the same perturbative
formalism developed in Sec.~II and are treated within the computational
framework introduced in Sec.~III.

\subsection{Reduction of the Detector Response}

The detector response for uniformly accelerated observers is obtained by
evaluating the Minkowski Wightman function along the Born-rigid
accelerated detector trajectories introduced in the previous subsection
and substituting the resulting correlation functions into the general
expressions for the excitation probabilities and exchange amplitude,
Eqs.~(\ref{eq:PGeneral}) and (\ref{eq:XGeneral}).

The local response depends only on the field correlation evaluated along
a single accelerated trajectory. As shown in Appendix~C, the
corresponding Wightman function has the thermal form associated with the
Unruh temperature of that detector. Since the two detectors have
different proper accelerations, their Unruh temperatures are
\begin{equation}
T_A=\frac{a}{2\pi},
\qquad
T_B=\frac{a}{2\pi(1+aL_R)}.
\end{equation}

Consequently, the local excitation probabilities are related separately
to the responses of inertial detectors immersed in thermal baths at
the corresponding Unruh temperatures. In terms of the dimensionless
parameters
\begin{equation}
\alpha=\Omega\sigma,
\qquad
\beta=\frac{L_R}{\sigma},
\qquad
\gamma=a\sigma,
\end{equation}
the responses are
\begin{equation}
P_A(\alpha,\gamma)
=
P_T\left(\alpha,\frac{\gamma}{2\pi}\right),
\end{equation}
and
\begin{equation}
P_B(\alpha,\gamma,\beta)
=
P_T\left(
\alpha,\frac{\gamma}{2\pi(1+\gamma\beta)}
\right).
\end{equation}
Here $P_T(\alpha,T\sigma)$ denotes the local response of an inertial
detector with the same energy gap and proper-time switching width in a
thermal field at temperature $T$. These relations were verified
independently through numerical evaluation of the corresponding
one-dimensional integrals.

The exchange amplitude, however, probes correlations between the two
Born-rigid accelerated detectors. Since the detectors occupy different
Rindler positions, they possess different proper accelerations and
evolve with different proper times. For Gaussian switching functions
of equal width in their respective proper times, the exchange amplitude
retains a genuinely two-dimensional integral and must be evaluated
directly using the accelerated trajectories and the Minkowski Wightman
function.

The complete derivation of both detector response functions and the
exchange amplitude is presented in Appendix~C. We denote the
corresponding excitation probabilities and exchange amplitude by
\begin{equation}
P_A=\mathcal{P}_A(\Omega,L_R,\sigma,a),
\end{equation}
\begin{equation}
P_B=\mathcal{P}_B(\Omega,L_R,\sigma,a),
\end{equation}
and
\begin{equation}
X_A=\mathcal{X}_A(\Omega,L_R,\sigma,a),
\end{equation}
respectively. Since the local responses are determined separately by
the two Unruh temperatures, they need not be equal. The exchange
amplitude depends on correlations between both accelerated trajectories
and cannot in general be obtained from the thermal response at either
detector's Unruh temperature. The harvesting analysis must therefore
account for both local excitation probabilities and the nonlocal
exchange amplitude.

\subsection{Harvesting Landscape for Uniformly Accelerated Detectors}

Having derived the detector response functions in
Appendix~\ref{app:Acceleration}, we now investigate the harvesting of
entanglement by uniformly accelerated detectors. As in the previous
sections, the detector configuration is characterized by the
dimensionless parameters
\begin{equation}
\alpha=\Omega\sigma,
\qquad
\beta=\frac{L_R}{\sigma},
\qquad
\gamma=a\sigma,
\end{equation}
where $\gamma$ measures the detector acceleration relative to the
interaction timescale. The harvesting problem is therefore specified
by the parameter space $(\alpha,\beta,\gamma)$.

The most immediate consequence of uniform acceleration appears in the
local detector responses. As demonstrated analytically in
Appendix~\ref{app:Acceleration}, the excitation probabilities satisfy
\begin{equation}
P_A(\alpha,\gamma)
=
P_T\left(\alpha,\frac{\gamma}{2\pi}\right),
\end{equation}
and
\begin{equation}
P_B(\alpha,\gamma,\beta)
=
P_T\left(
\alpha,\frac{\gamma}{2\pi(1+\gamma\beta)}
\right).
\end{equation}
These are the responses of inertial detectors immersed in thermal baths
at their respective Unruh temperatures. The two responses must be
evaluated separately because the proper accelerations, and hence the
Unruh temperatures, are different.

\begin{figure}
    \centering
    \includegraphics[width=0.52\linewidth]{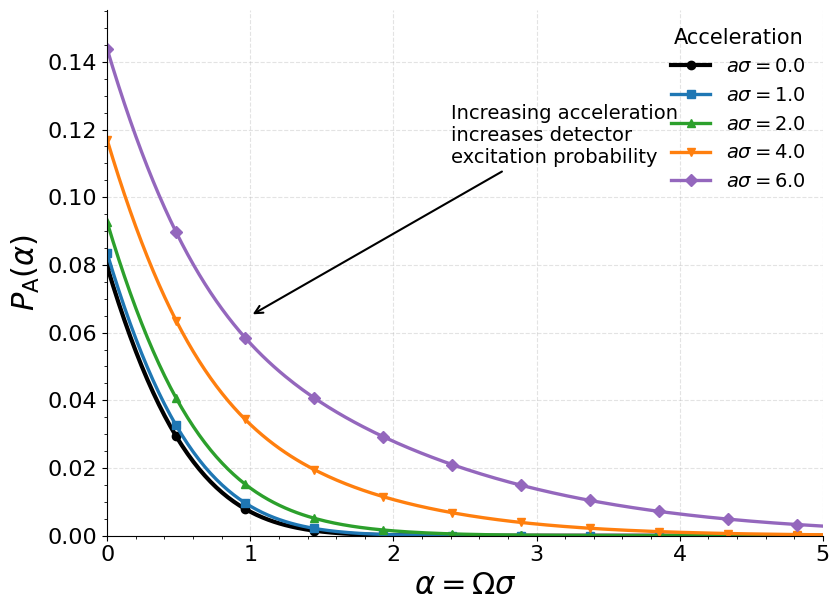}
    \caption{
Local detector excitation probability $P_A(\alpha)$ for detector $A$
under Born-rigid acceleration, as a function of the dimensionless
detector gap $\alpha=\Omega\sigma$ for representative values of the
acceleration parameter $\gamma=a\sigma$. Increasing acceleration
produces larger excitation probabilities, consistent with the
Unruh-induced thermal response derived in Appendix~C.
}
    \label{fig:AccelerationP}
\end{figure}

Figure~\ref{fig:AccelerationP} illustrates the local response of
detector $A$. Increasing acceleration enhances its excitation
probability, reflecting the increasing thermal character of the vacuum
perceived by this detector.

The situation is markedly different for the exchange amplitude.
Although each local response has a thermal analogue, the exchange
amplitude depends on correlations between two detectors located at
different Rindler positions. Since the detectors experience different
proper accelerations and evolve according to different proper times, the
two-dimensional integral describing $X_A$ cannot in general be reduced
to the thermal expression. Consequently, the acceleration dependence of
the nonlocal correlations must be computed independently.

\begin{figure}
    \centering
    \includegraphics[width=0.52\linewidth]{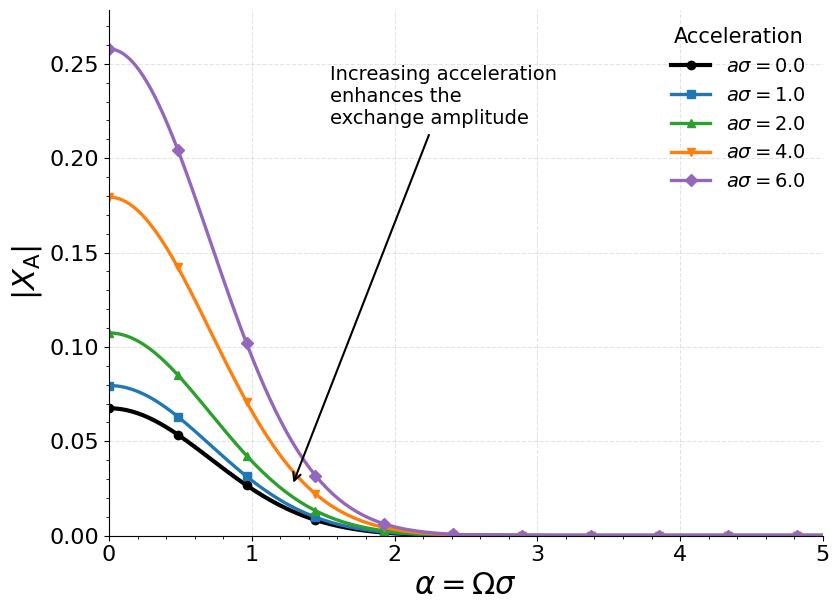}
    \caption{
Exchange amplitude $|X_A|$ as a function of the dimensionless detector
gap $\alpha=\Omega\sigma$ for representative accelerations
$a\sigma$. Although the local detector responses are thermally
equivalent through the Unruh effect at their respective temperatures,
the exchange amplitude depends explicitly on the geometry of the
accelerated detector trajectories and therefore exhibits
acceleration-dependent behaviour.
}
    \label{fig:AccelerationX}
\end{figure}

Figure~\ref{fig:AccelerationX} shows the resulting exchange amplitude.
Unlike the local response, its behaviour reflects the geometry of the
accelerated detector trajectories rather than a simple thermal
replacement. The nonlocal exchange amplitude exhibits a clear enhancement with acceleration at small detector gaps, although this ordering is not maintained throughout the entire range of \(\alpha\). This indicates that acceleration modifies not only the local detector response but also the nonlocal vacuum correlations available for harvesting.

The reduced detector density matrix depends on the two local excitation
probabilities $P_A$ and $P_B$, together with the exchange amplitude
$X_A$. For unequal local responses, the leading-order entanglement
negativity is
\begin{equation}
\mathcal{N}_A
=
\max\left\{
0,
\frac{
\sqrt{(P_A-P_B)^2+4|X_A|^2}
-(P_A+P_B)
}{2}
\right\}.
\label{eq:AccelerationNegativityGeneral}
\end{equation}
Thus, entanglement harvesting occurs when the nonlocal exchange
correlation satisfies
\begin{equation}
|X_A|^2>P_AP_B.
\label{eq:AccelerationHarvestCondition}
\end{equation}
The familiar condition $|X_A|>P_A$ follows only in the special case
$P_A=P_B$ and cannot in general be used for the Born-rigid configuration
considered here.

\begin{figure}
    \centering
    \includegraphics[width=0.52\linewidth]{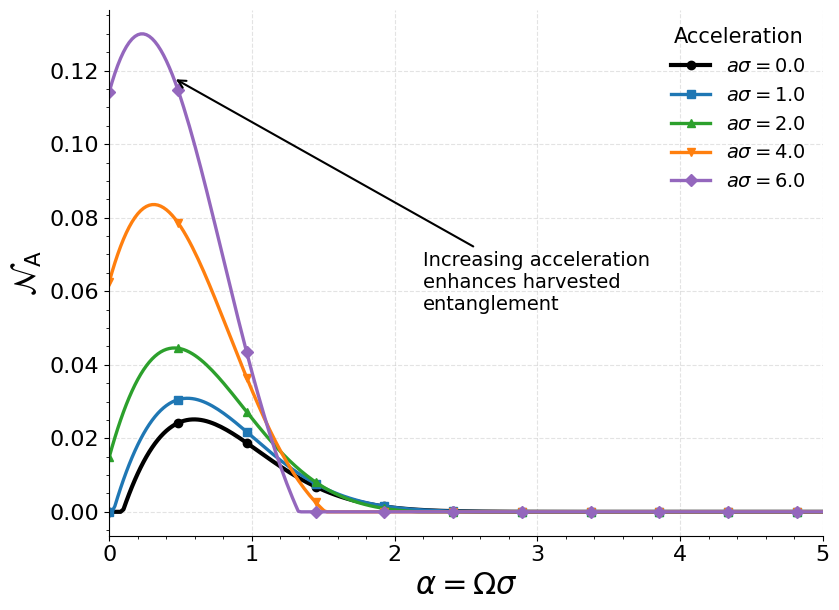}
    \caption{
Harvested entanglement negativity $\mathcal{N}_A$ as a function of the
dimensionless detector gap $\alpha=\Omega\sigma$ for different values of
the acceleration parameter $a\sigma$, with fixed detector separation
$\beta=L_R/\sigma=1$. The negativity is calculated using the
unequal-response expression in
Eq.~(\ref{eq:AccelerationNegativityGeneral}), with the local excitation
probabilities of both detectors evaluated separately.
}
    \label{fig:AccelerationNegativity}
\end{figure}

Figure~\ref{fig:AccelerationNegativity} presents the harvested
entanglement for a representative detector separation. The negativity
can display a maximum as a function of the detector gap, reflecting
the competition between local detector excitations and nonlocal field
correlations. Its quantitative dependence on acceleration must be
determined using Eq.~(\ref{eq:AccelerationNegativityGeneral}), with
both local responses evaluated separately.

\begin{figure}
    \centering
    \includegraphics[width=0.55\linewidth]{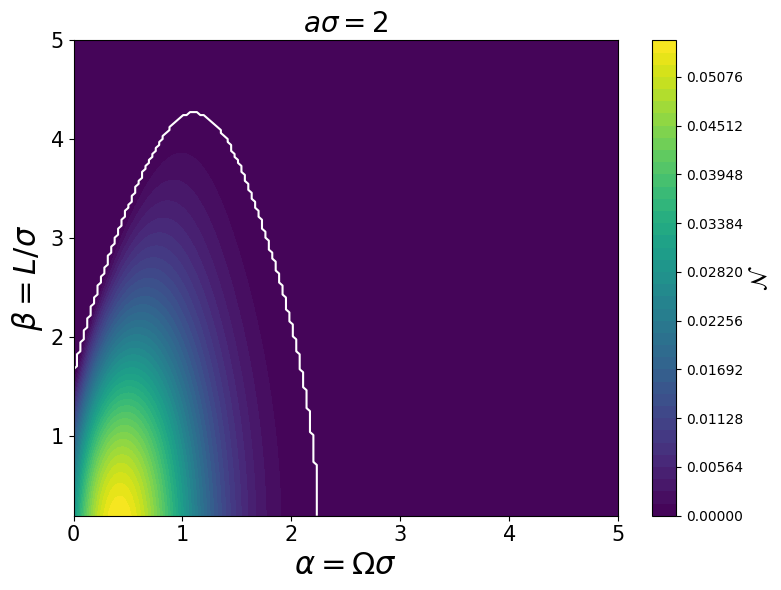}
    \caption{
Harvesting landscape for two uniformly accelerated
Unruh--DeWitt detectors with representative acceleration
$a\sigma=2$. The colour map shows the entanglement negativity
$\mathcal{N}_A$ over the $(\alpha,\beta)$ parameter space, while the
white contour denotes the harvesting boundary $\mathcal{N}_A=0$,
equivalently $|X_A|^2=P_AP_B$. Although the local responses have
thermal analogues at their respective Unruh temperatures, the harvesting
boundary also depends on the exchange amplitude and the unequal local
excitation probabilities.
}
    \label{fig:AccelerationLandscape}
\end{figure}

Finally, the complete harvesting landscape is obtained by repeating the
calculation over the $(\alpha,\beta)$ parameter space for fixed
acceleration. Figure~\ref{fig:AccelerationLandscape} shows the resulting
entanglement landscape for the representative case
$a\sigma=2$. The colour scale represents the harvested negativity,
whereas the white contour marks the boundary $\mathcal{N}_A=0$,
separating the entangled and separable regions.

Comparison with the corresponding thermal landscape reveals an
interesting distinction. Although the local responses have thermal
analogues through the Unruh correspondence, the harvesting boundary
need not coincide with the thermal result. The difference originates
from the exchange amplitude and from the unequal local responses in
the Born-rigid configuration, demonstrating that the Unruh equivalence
for local detector excitations does not fully reproduce the nonlocal
vacuum correlations responsible for entanglement harvesting.

\section{Conclusion and Discussion}

Entanglement harvesting provides an operational window into the nonlocal correlation structure of quantum fields through the local interaction of spatially separated quantum detectors. While the underlying perturbative formalism has been applied to a wide variety of spacetime backgrounds and quantum field states, existing studies have largely focused on individual physical scenarios and specific detector configurations. One of the primary motivations of the present work has therefore been to organize these seemingly different harvesting problems within a common theoretical and computational framework and to use this framework to explore the resulting multidimensional parameter-space structure.

Beginning from the general interaction between localized Unruh--DeWitt detectors and a scalar quantum field, we derived the reduced two-detector density matrix in a geometry-independent form. At leading nontrivial order in perturbation theory, the detector state is determined by the local excitation probabilities and the nonlocal exchange amplitude, with the dependence on the quantum field state and spacetime entering through the corresponding Wightman two-point function. This separation provides a common mathematical basis for comparing different detector trajectories and field states. The detector dynamics and entanglement criterion remain the same across the configurations considered here, while the physical setting is incorporated through the detector trajectories and the associated Wightman function.

A central technical component of the work is the analytical reduction of the detector response. For configurations in which the required stationarity and time-translation properties are present, the oscillatory double integrals can be reduced exactly to one-dimensional integral representations by introducing relative and average proper-time coordinates and exploiting the Gaussian switching profile. This reduction substantially improves the efficiency of numerical evaluation and makes systematic parameter-space scans feasible. The accelerated configuration also illustrates an important limitation of this reduction: while the local response admits the same one-dimensional treatment, the nonlocal exchange amplitude retains a two-dimensional integral because the two Born-rigid accelerated trajectories have different proper-time parametrizations. Thus, the common framework should be understood as a shared detector-level formalism and computational pipeline, rather than as a requirement that every physical configuration reduce to exactly the same integral dimensionality.

The use of dimensionless parameters provides a further advantage. The detector energy gap, spatial separation, temperature, acceleration, and interaction timescale can be expressed through dimensionless combinations such as $\Omega\sigma$, $L/\sigma$, $T\sigma$, and $a\sigma$. This allows the harvesting response to be represented as a landscape over parameter space rather than through isolated values or one-dimensional scans. Such landscapes make it possible to identify harvesting regions, phase boundaries, optimal parameter ranges, and the relative importance of local excitation noise and nonlocal detector correlations.

The inertial detectors in the Minkowski vacuum provide the natural benchmark for this analysis. In this case, the Wightman function has a simple translationally invariant form and the detector trajectories are stationary, allowing both the local excitation probability and the nonlocal exchange amplitude to be reduced analytically to one-dimensional integrals. The resulting numerical implementation reproduces the corresponding analytical benchmark and establishes the baseline harvesting landscape against which the other configurations can be compared. The Minkowski analysis also makes explicit the characteristic competition between detector separation and energy gap: increasing separation suppresses the nonlocal correlations, while increasing the detector gap suppresses the detector response and modifies the region in which the exchange contribution can overcome local excitation noise.

For thermal quantum fields, we found that increasing temperature enhances the local detector response while generally suppressing the magnitude of the nonlocal exchange amplitude. Within the parameter range investigated and for the perturbative Gaussian-switching framework considered here, these competing effects lead to a suppression of harvested entanglement, with the harvesting region becoming progressively smaller as the temperature increases. This behaviour illustrates how finite temperature modifies both local detector excitations and nonlocal field correlations, with the net effect determined by their relative contributions to the detector state. In the parameter range explored, this competition results in a reduction of the region satisfying the harvesting condition, rather than an enhancement of harvested entanglement.

The accelerated-detector analysis reveals a complementary and particularly important feature of the problem. Uniformly accelerated detectors in the Minkowski vacuum exhibit the Unruh effect in their local responses, which can be related to thermal responses at the corresponding Unruh temperatures. For the Born-rigid configuration considered here, however, the two detectors have different proper accelerations and consequently different Unruh temperatures. Their local excitation probabilities are therefore generally unequal and must be evaluated separately when determining whether entanglement is harvested.

This local thermal correspondence does not extend in general to the nonlocal exchange amplitude. The two detectors occupy different positions in the accelerated frame, and their different proper-time parametrizations enter the joint correlation function in a way that cannot be reproduced simply by replacing the Minkowski vacuum with a thermal state. Consequently, the harvesting landscape of the accelerated configuration is not determined solely by the local Unruh temperatures. The nonlocal correlations between the two trajectories must also be taken into account, and the entanglement criterion must accommodate both local excitation probabilities. This distinction demonstrates that local thermal behaviour alone is insufficient to characterize bipartite entanglement harvesting.

For the accelerated configuration investigated here, increasing the acceleration modifies both the local responses and the nonlocal exchange contribution. The resulting harvesting behaviour reflects the competition between these contributions, with the two local excitation probabilities evaluated separately for the Born-rigid detectors. More generally, the accelerated results show that observer motion can reshape the harvesting landscape in a manner that cannot be inferred solely from the local Unruh temperatures. This provides a useful example of how entanglement harvesting can distinguish between physical situations that may appear equivalent from the perspective of a single local detector.

Taken together, the three configurations considered in this work demonstrate the usefulness of viewing entanglement harvesting as a multidimensional landscape. Rather than asking only whether two detectors harvest entanglement for a particular choice of parameters, the landscape approach reveals how the harvesting region evolves as detector separation, energy gap, interaction duration, temperature, and acceleration are varied. It also provides a direct way of identifying phase boundaries and comparing the relative sensitivity of local and nonlocal detector responses. In this sense, the landscape representation complements the conventional analysis of individual detector configurations by providing a global view of the regimes in which quantum correlations can be extracted.

The comparison between thermal and accelerated configurations is especially instructive in this regard. Although the Unruh effect establishes a thermal correspondence at the level of the local response of an accelerated detector, Born-rigid acceleration gives the two detectors different proper accelerations and Unruh temperatures. Moreover, the nonlocal harvesting behaviour retains information about the joint spacetime trajectories and cannot be inferred from the local responses alone. The results therefore support a distinction between \emph{local thermal equivalence} and \emph{nonlocal harvesting equivalence}: the former may hold separately for each detector under the appropriate identification of temperature, while the latter does not follow from local equivalence alone. This observation reinforces the potential of entanglement harvesting as an operational probe of quantum field correlations beyond what can be learned from single-detector observables.

The framework developed here also suggests several directions for future work. The present study is restricted to identical two-level Unruh--DeWitt detectors coupled to a massless scalar field and treated to leading nontrivial order in the detector--field coupling. A natural extension is to investigate massive fields, different switching profiles, nonidentical detector gaps, and more general detector trajectories. The same detector-level organization can also be used as a starting point for studying quantum fields in curved spacetime, cosmological backgrounds, black-hole geometries, and anti-de~Sitter spacetime, with the corresponding geometric and state-dependent information entering through the appropriate field correlation functions and detector kinematics. Extensions to electromagnetic fields, squeezed states, and other non-vacuum field states would provide further tests of the robustness of the landscape approach.

Furthermore, going beyond leading-order perturbation theory would allow the regime of stronger detector--field coupling to be explored and would test the extent to which the leading-order harvesting criterion remains reliable. Increasing the number of detectors would open the possibility of studying multipartite harvesting and the structure of distributed quantum correlations, while time-dependent detector networks could provide a setting for investigating relativistic quantum communication. More generally, combining harvesting landscapes with experimentally motivated detector models may provide a route toward using localized quantum probes to reconstruct features of spacetime and field correlations that are otherwise inaccessible to local measurements.

In summary, the present work develops a common theoretical and computational organization for studying entanglement harvesting across inertial, thermal, and uniformly accelerated configurations. The analysis shows how analytical reduction and dimensionless parameterization make systematic exploration of harvesting landscapes possible, while the comparison of thermal and accelerated detectors demonstrates that correspondence between local responses does not necessarily imply equivalent nonlocal harvesting behaviour. In particular, the Born-rigid configuration highlights the importance of accounting for unequal local excitation probabilities and trajectory-dependent correlations when assessing harvested entanglement. We therefore expect the landscape-based approach developed here to provide a useful basis for future investigations of quantum correlations in more general relativistic quantum-field settings.

\section{Acknowledgments}
R.N and A.K thank their institute, BITS Pilani Hyderabad campus, for providing the required infrastructure for this research work.
\appendix

\section{Distributional Reduction of Detector Response Integrals}
\label{app:distribution}

In this appendix we develop a general computational framework for evaluating
the local detector response for smooth switching functions.
The derivation is independent of the particular spacetime geometry and relies
only on the distributional structure of the Wightman function.
The resulting reduction will therefore serve as the common analytical
foundation for all spacetime backgrounds considered in this work.


\subsection{Distributional reduction}

The one-dimensional detector response integral derived in Sec.~III contains
a second-order pole, which is most conveniently treated in the sense of
tempered distributions. Rather than evaluating the singular integral
directly, we first reduce it to an equivalent principal-value integral
involving only a simple pole.

The Sokhotski--Plemelj formula reads

\begin{equation}
\frac{1}{u-i0}
=
\operatorname{PV}\!\left(\frac1u\right)
+i\pi\delta(u),
\label{eq:A9}
\end{equation}

and the second-order distribution is defined recursively through

\begin{equation}
\frac{1}{(u-i0)^2}
=
-
\frac{d}{du}
\left(
\frac{1}{u-i0}
\right).
\label{eq:A10}
\end{equation}

Let $f(u)$ be a Schwartz-class test function.
Using the definition of the distributional derivative,

\[
\langle T',f\rangle
=
-
\langle T,f'\rangle,
\]

together with Eq.~(\ref{eq:A9}), one obtains

\begin{equation}
\boxed{
\int_{-\infty}^{\infty}
\frac{f(u)}
{(u-i0)^2}
\,du
=
\operatorname{PV}
\int_{-\infty}^{\infty}
\frac{f'(u)}{u}
\,du
+
i\pi f'(0).
}
\label{eq:A11}
\end{equation}

Equation~(\ref{eq:A11}) is the key analytical result of this appendix.
It replaces the second-order singularity by a principal-value integral
containing only a simple pole together with a local distributional
contribution arising from the derivative of the propagator.


\subsection{Application to Gaussian switching}

For the Gaussian switching function,

\begin{equation}
f(u)
=
\exp\!\left(
-\frac{u^{2}}
{4\sigma^{2}}
\right)
e^{-i\Omega u},
\label{eq:A12}
\end{equation}

the derivative is

\begin{equation}
f'(u)
=
\left(
-\frac{u}{2\sigma^{2}}
-i\Omega
\right)
f(u).
\label{eq:A13}
\end{equation}

Substituting Eq.~(\ref{eq:A13}) into the distributional identity
(\ref{eq:A11}) yields

\begin{equation}
P
=
-
\frac{\lambda^{2}\sigma}
{4\pi^{3/2}}
\left[
\operatorname{PV}
\int_{-\infty}^{\infty}
\frac{f'(u)}u\,du
+
i\pi f'(0)
\right].
\label{eq:A14}
\end{equation}

The problem has therefore been reduced to evaluating a single
principal-value integral together with the local contribution
$f'(0)=-i\Omega$, both of which admit closed analytical expressions.


\subsection{Evaluation of the principal-value integral}

Using Eq.~(\ref{eq:A13}), the principal-value integral becomes

\begin{equation}
\operatorname{PV}
\int
\frac{f'(u)}u\,du
=
-
\frac{1}{2\sigma^{2}}
\int
f(u)\,du
-
i\Omega
\operatorname{PV}
\int
\frac{f(u)}u\,du.
\label{eq:A15}
\end{equation}

The first integral is the Fourier transform of a Gaussian,

\begin{equation}
\int_{-\infty}^{\infty}
e^{-u^{2}/4\sigma^{2}}
e^{-i\Omega u}
\,du
=
2\sigma\sqrt{\pi}
\,e^{-\Omega^{2}\sigma^{2}}.
\label{eq:A16}
\end{equation}

Introducing the dimensionless parameter

\[
\alpha=\Omega\sigma,
\]

the remaining principal-value integral may be written as

\begin{equation}
I(\alpha)
=
\operatorname{PV}
\int_{-\infty}^{\infty}
\frac{
e^{-x^{2}}
e^{-2i\alpha x}
}{x}
\,dx.
\label{eq:A17}
\end{equation}

Rather than evaluating Eq.~(\ref{eq:A17}) directly, it is convenient to
differentiate with respect to the parameter $\alpha$. Since the resulting
integral is absolutely convergent,

\begin{equation}
\frac{dI}{d\alpha}
=
-2i
\int_{-\infty}^{\infty}
e^{-x^{2}}
e^{-2i\alpha x}
\,dx
=
-2i\sqrt{\pi}\,
e^{-\alpha^{2}}.
\label{eq:A18}
\end{equation}

Integrating Eq.~(\ref{eq:A18}) with respect to $\alpha$
introduces an integration constant.
This constant is determined by evaluating Eq.~(\ref{eq:A17}) at
$\alpha=0$, for which the integrand is odd and the principal-value
integral therefore vanishes,
\[
I(0)=0.
\]
One therefore obtains

\begin{equation}
\boxed{
I(\alpha)
=
-i\pi\,
\operatorname{erf}(\alpha).
}
\label{eq:A19}
\end{equation}

Substituting Eqs.~(\ref{eq:A16}) and (\ref{eq:A19}) into
Eq.~(\ref{eq:A15}), and subsequently into
Eq.~(\ref{eq:A14}), yields the closed-form detector response in
Minkowski spacetime. More importantly, the derivation above
demonstrates that the original oscillatory integral may be reduced
systematically to elementary Gaussian integrals together with a
single special function. This analytical reduction provides the exact
benchmark against which the numerical implementation developed in
Appendix~B is validated.

\subsection{Closed-form detector response}

Combining Eqs.~(\ref{eq:A15}), (\ref{eq:A16}) and (\ref{eq:A19}), and using

\[
1-\operatorname{erf}(\alpha)
=
\operatorname{erfc}(\alpha),
\]

the detector response assumes the compact analytical form

\begin{equation}
\boxed{
P(\alpha)
=
\frac{\lambda^{2}}{4\pi}
\left[
e^{-\alpha^{2}}
-
\sqrt{\pi}\,
\alpha\,
\operatorname{erfc}(\alpha)
\right].
}
\label{eq:A20}
\end{equation}

Equation~(\ref{eq:A20}) is positive for all detector gaps and decreases monotonically with increasing $\alpha$, consistent with the expected suppression of detector excitation at large energy gaps. This closed-form expression provides an exact analytical benchmark
for the numerical evaluation of the one-dimensional detector response
integrals presented in Appendix~B, where the corresponding one-dimensional integral representation is evaluated directly.


\section{Reduction of the Detector Response and Exchange Amplitude}
\label{app:Reduction}

In this appendix we derive the one-dimensional integral representations
used throughout the numerical calculations in Secs.~IV--VI.
These expressions are obtained directly from the general detector response
derived in Sec.~II and form the basis of the numerical implementation.
For the Minkowski vacuum, both the detector response and exchange amplitude
may furthermore be expressed in closed analytical form, providing an
independent benchmark for the numerical calculations.


\subsection{Local Detector Response}

For inertial detectors in Minkowski spacetime the excitation probability is

\begin{equation}
P
=
\lambda^2
\int_{-\infty}^{\infty}
d\tau
\int_{-\infty}^{\infty}
d\tau'
\,
\chi(\tau)
\chi(\tau')
e^{-i\Omega(\tau-\tau')}
W_M^{\mathrm{loc}}(\tau,\tau'),
\label{eq:B1}
\end{equation}

where

\begin{equation}
W_M^{\mathrm{loc}}
=
-
\frac{1}{4\pi^2}
\frac{1}
{(\tau-\tau'-i\epsilon)^2}.
\label{eq:B2}
\end{equation}

Introducing

\begin{equation}
u=\tau-\tau',
\qquad
v=\frac{\tau+\tau'}{2},
\label{eq:B3}
\end{equation}

whose Jacobian is unity,

\[
d\tau\,d\tau'=du\,dv,
\]

the Gaussian switching functions become

\begin{equation}
\chi(\tau)\chi(\tau')
=
\exp\!\left[
-\frac{v^2}{\sigma^2}
-
\frac{u^2}{4\sigma^2}
\right].
\label{eq:B4}
\end{equation}

The integral over the average time evaluates immediately,

\[
\int_{-\infty}^{\infty}
e^{-v^2/\sigma^2}
dv
=
\sqrt{\pi}\sigma,
\]

yielding

\begin{equation}
P
=
-
\frac{\lambda^2\sigma}
{4\pi^{3/2}}
\int_{-\infty}^{\infty}
du
\,
\frac{
e^{-u^2/(4\sigma^2)}
e^{-i\Omega u}
}
{(u-i\epsilon)^2}.
\label{eq:B5}
\end{equation}

Equation~(\ref{eq:B5}) is precisely the one-dimensional representation
whose distributional evaluation is presented in Appendix~A and which
forms the basis of the numerical implementation.


\subsection{Exchange Amplitude}

The exchange amplitude is

\begin{equation}
X
=
-\lambda^2
\int d\tau
\int d\tau'
\,
\chi(\tau)
\chi(\tau')
e^{+i\Omega(\tau+\tau')}
W_M^{\mathrm{nonloc}}(\tau,\tau'),
\label{eq:B6}
\end{equation}

in agreement with Eq.~(\ref{eq:XGeneral}) of the main text, where

\begin{equation}
W_M^{\mathrm{nonloc}}
=
-
\frac{1}{4\pi^2}
\frac{1}
{(\tau-\tau'-i\epsilon)^2-L^2}.
\label{eq:B7}
\end{equation}

Using the variables introduced in Eq.~(\ref{eq:B3}),

\[
e^{+i\Omega(\tau+\tau')}
=
e^{2i\Omega v},
\]

while the Gaussian factorization remains unchanged.

The average-time integration gives

\begin{equation}
\int_{-\infty}^{\infty}
dv\,
e^{-v^2/\sigma^2}
e^{2i\Omega v}
=
\sqrt{\pi}\sigma
e^{-\Omega^2\sigma^2},
\label{eq:B8}
\end{equation}

so that

\begin{equation}
X
=
\frac{\lambda^2\sigma}
{4\pi^{3/2}}
e^{-\Omega^2\sigma^2}
\int_{-\infty}^{\infty}
du
\,
\frac{
e^{-u^2/(4\sigma^2)}
}
{(u-i\epsilon)^2-L^2}.
\label{eq:B9}
\end{equation}

Introducing

\begin{equation}
\alpha=\Omega\sigma,
\qquad
\beta=\frac{L}{\sigma},
\qquad
x=\frac{u}{2\sigma},
\label{eq:B10}
\end{equation}

one obtains

\begin{equation}
X(\alpha,\beta)
=
\frac{\lambda^2}{2\pi^{3/2}}
e^{-\alpha^2}
\operatorname{PV}
\int_{-\infty}^{\infty}
dx\,
\frac{e^{-x^2}}{4x^2-\beta^2}.
\label{eq:B11}
\end{equation}

The principal-value integral in Eq.~(\ref{eq:B11}) may be evaluated
analytically by first expressing the denominator through a partial-fraction
decomposition and then applying the standard Hilbert transform identity for
a Gaussian. One obtains

\begin{equation}
\operatorname{PV}
\int_{-\infty}^{\infty}
dx\,
\frac{e^{-x^{2}}}
{4x^{2}-\beta^{2}}
=
-
\frac{\sqrt{\pi}}{\beta}
D\!\left(\frac{\beta}{2}\right),
\label{eq:B12}
\end{equation}

where

\begin{equation}
D(z)
=
e^{-z^{2}}
\int_{0}^{z}
e^{t^{2}}\,dt
\end{equation}

is Dawson's integral.

Substituting Eq.~(\ref{eq:B12}) into Eq.~(\ref{eq:B11}) yields the
closed-form exchange amplitude

\begin{equation}
\boxed{
X(\alpha,\beta)
=-
\frac{\lambda^{2}}
{2\pi}
\,
\frac{e^{-\alpha^{2}}}{\beta}
\,
D\!\left(\frac{\beta}{2}\right).
}
\label{eq:B13}
\end{equation}
Equation~(\ref{eq:B13}) provides the analytical benchmark used to validate
the numerical implementation.


\subsection{Numerical Validation}

To verify the analytical reduction, the principal-value integral in
Eq.~(\ref{eq:B11}) was evaluated independently by direct numerical
integration and compared with the analytical expression
Eq.~(\ref{eq:B12}). Representative values are listed in
Table~\ref{tab:validation}. The agreement is seen to be at the level of the
numerical integration accuracy.

\begin{table}[h]
\centering
\begin{tabular}{cccc}
\hline
$\beta$ &
Numerical &
Analytical &
Absolute error \\
\hline
0.5 &
$-0.8502030010$ &
$-0.8502076981$ &
$4.70\times10^{-6}$ \\

1.0 &
$-0.7522923448$ &
$-0.7522939024$ &
$1.56\times10^{-6}$ \\

1.5 &
$-0.6180098398$ &
$-0.6180106628$ &
$8.23\times10^{-7}$ \\

2.0 &
$-0.4768600872$ &
$-0.4768605471$ &
$4.60\times10^{-7}$ \\

3.0 &
$-0.2530171213$ &
$-0.2530172384$ &
$1.17\times10^{-7}$ \\

4.0 &
$-0.1335279637$ &
$-0.1335279832$ &
$1.95\times10^{-8}$ \\
\hline
\end{tabular}
\caption{Comparison between the direct numerical evaluation of the
principal-value integral in Eq.~(\ref{eq:B11}) and the analytical result,
Eq.~(\ref{eq:B12}).}
\label{tab:validation}
\end{table}

The excellent agreement confirms the correctness of the analytical
evaluation of the principal-value integral and validates the numerical
implementation employed throughout the harvesting calculations presented in
the main text.


\subsection{Thermal Detector Response}

For a massless scalar field in a thermal state of temperature $T$, the
Wightman function is

\begin{equation}
W_T^{\text{local}}(u)
=
-
\frac{(\pi T)^2}{4\pi^2}
\frac{1}
{\sinh^2\!\left[\pi T(u-i\epsilon)\right]},
\label{eq:BT1}
\end{equation}

where $u=\tau-\tau'$ denotes the relative detector proper time.
Introducing the relative and average time coordinates,

\[
u=\tau-\tau',
\qquad
v=\frac{\tau+\tau'}{2},
\]

the Gaussian switching functions factorize exactly as in the preceding
subsection, allowing the integration over the average interaction
time to be carried out analytically. One obtains

\begin{equation}
P_T
=
-
\frac{\lambda^2\sigma(\pi T)^2}
{4\pi^{3/2}}
\int_{-\infty}^{\infty}
du\,
\frac{
e^{-u^2/(4\sigma^2)}
e^{-i\Omega u}
}
{\sinh^2\!\left[\pi T(u-i\epsilon)\right]}.
\label{eq:BT2}
\end{equation}

Introducing the dimensionless variables

\begin{equation}
\alpha=\Omega\sigma,
\qquad
\theta=T\sigma,
\qquad
x=\frac{u}{2\sigma},
\label{eq:BT3}
\end{equation}

reduces Eq.~(\ref{eq:BT2}) to

\begin{equation}
P_T
=
-
\frac{\lambda^2\sqrt{\pi}\,\theta^2}
{2}
\int_{-\infty}^{\infty}
dx\,
\frac{
e^{-x^2}
e^{-2i\alpha x}
}
{\sinh^2\!\left[2\pi\theta
\left(x-i\epsilon/(2\sigma)\right)\right]}.
\label{eq:BT4}
\end{equation}

As in the Minkowski case, the thermal Wightman function possesses the
same short-distance singularity as the vacuum correlation function.
Expanding the denominator for small $x$,

\begin{equation}
\frac{\theta^2}
{\sinh^2(2\pi\theta x)}
=
\frac{1}{4\pi^2x^2}
-
\frac{\theta^2}{3}
+
\mathcal{O}(x^2),
\label{eq:BT5}
\end{equation}

shows that the ultraviolet divergence is identical to that of the
Minkowski vacuum. It is therefore convenient to isolate this known
contribution by writing

\begin{equation}
\frac{\theta^2}
{\sinh^2(2\pi\theta x)}
=
\frac{1}{4\pi^2x^2}
+
\left[
\frac{\theta^2}
{\sinh^2(2\pi\theta x)}
-
\frac{1}{4\pi^2x^2}
\right].
\label{eq:BT6}
\end{equation}

The first term reproduces exactly the Minkowski detector response
derived in Appendix~A, while the expression enclosed in brackets is
finite at $x=0$,

\[
\frac{\theta^2}
{\sinh^2(2\pi\theta x)}
-
\frac{1}{4\pi^2x^2}
=
-\frac{\theta^2}{3}
+
\mathcal{O}(x^2),
\]

and therefore contains no ultraviolet singularity. The detector
response may consequently be written as

\begin{equation}
P_T(\alpha,\theta)
=
P_M(\alpha)
+
\Delta P_T(\alpha,\theta),
\label{eq:BT7}
\end{equation}

where the thermal correction is

\begin{equation}
\Delta P_T
=
-
\frac{\lambda^2\sqrt{\pi}}
{2}
\int_{-\infty}^{\infty}
dx\,
e^{-x^2}
e^{-2i\alpha x}
\left[
\frac{\theta^2}
{\sinh^2(2\pi\theta x)}
-
\frac{1}{4\pi^2x^2}
\right].
\label{eq:BT8}
\end{equation}

The subtracted kernel in Eq.~(\ref{eq:BT8}) is finite over the entire
integration domain. Although the kernel itself decays algebraically at
large $|x|$, the Gaussian switching factor ensures rapid decay of the
full integrand. Consequently, the thermal correction may be evaluated
efficiently using standard adaptive quadrature routines without requiring
principal-value prescriptions or distributional identities for the
subtracted integral. The complete detector response is obtained by
adding the exact Minkowski contribution, Eq.~(\ref{eq:A20}), to the
numerically evaluated thermal correction. This formulation provides a
stable numerical implementation over the full parameter range considered
in the present work while automatically recovering the zero-temperature
limit,

\[
\lim_{\theta\rightarrow0}
P_T(\alpha,\theta)
=
P_M(\alpha).
\]


\subsection{Thermal Exchange Amplitude}

We now derive the one-dimensional integral representation of the
exchange amplitude for detectors interacting with a massless scalar
field in a thermal state. The nonlocal thermal Wightman function
for two inertial detectors separated by a distance $L$ is

\begin{equation}
W_T^{\mathrm{nonloc}}(u,L)
=
-\frac{T}{8\pi L}
\left[
\coth\!\left(\pi T(u-L-i\epsilon)\right)
-
\coth\!\left(\pi T(u+L-i\epsilon)\right)
\right],
\label{eq:BTX1}
\end{equation}

where $u=\tau-\tau'$ is the relative proper time.

The exchange amplitude is given by

\begin{equation}
X_T
=
-\lambda^2
\int_{-\infty}^{\infty}d\tau
\int_{-\infty}^{\infty}d\tau'
\,
\chi(\tau)\chi(\tau')
e^{i\Omega(\tau+\tau')}
W_T^{\mathrm{nonloc}}(\tau-\tau',L).
\label{eq:BTX2}
\end{equation}

Introducing the relative and average time coordinates,

\begin{equation}
u=\tau-\tau',
\qquad
v=\frac{\tau+\tau'}{2},
\label{eq:BTX3}
\end{equation}

with unit Jacobian, the Gaussian switching functions factorize as

\begin{equation}
\chi(\tau)\chi(\tau')
=
\exp\!\left[
-\frac{v^2}{\sigma^2}
-\frac{u^2}{4\sigma^2}
\right].
\label{eq:BTX4}
\end{equation}

Furthermore, the detector phase factor becomes

\begin{equation}
e^{i\Omega(\tau+\tau')}
=
e^{2i\Omega v}.
\end{equation}

The integration over the average time can therefore be performed
analytically:

\begin{equation}
\int_{-\infty}^{\infty}
dv\,
e^{-v^2/\sigma^2}
e^{2i\Omega v}
=
\sqrt{\pi}\sigma
e^{-\Omega^2\sigma^2}.
\label{eq:BTX5}
\end{equation}

Substituting this result into Eq.~(\ref{eq:BTX2}), we obtain

\begin{equation}
X_T
=
-\lambda^2\sqrt{\pi}\sigma
e^{-\Omega^2\sigma^2}
\int_{-\infty}^{\infty}
du\,
e^{-u^2/(4\sigma^2)}
W_T^{\mathrm{nonloc}}(u,L).
\label{eq:BTX6}
\end{equation}

Using the thermal Wightman function from
Eq.~(\ref{eq:BTX1}), this becomes

\begin{equation}
\begin{aligned}
X_T
={}&
\frac{\lambda^2\sqrt{\pi}\sigma T}
{8\pi L}
e^{-\Omega^2\sigma^2}
\\
&\times
\int_{-\infty}^{\infty}du\,
e^{-u^2/(4\sigma^2)}
\left[
\coth\!\left(\pi T(u-L-i\epsilon)\right)
\right.
\\
&\hspace{3.5cm}\left.
-
\coth\!\left(\pi T(u+L-i\epsilon)\right)
\right].
\end{aligned}
\label{eq:BTX7}
\end{equation}

Introducing the dimensionless variables

\begin{equation}
\alpha=\Omega\sigma,
\qquad
\beta=\frac{L}{\sigma},
\qquad
\theta=T\sigma,
\qquad
x=\frac{u}{2\sigma},
\label{eq:BTX8}
\end{equation}

and defining the dimensionless regulator
$\delta=\epsilon/(2\sigma)$, Eq.~(\ref{eq:BTX7}) reduces to

\begin{equation}
\begin{aligned}
X_T(\alpha,\beta,\theta)
={}&
\frac{\lambda^2\theta}
{4\sqrt{\pi}\,\beta}
e^{-\alpha^2}
\\
&\times
\int_{-\infty}^{\infty}dx\,
e^{-x^2}
\left[
\coth\!\left(
2\pi\theta
\left(x-\frac{\beta}{2}-i\delta\right)
\right)
\right.
\\
&\hspace{3.2cm}\left.
-
\coth\!\left(
2\pi\theta
\left(x+\frac{\beta}{2}-i\delta\right)
\right)
\right]
\end{aligned}
\label{eq:BTX9}
\end{equation}
The integral in Eq.~(\ref{eq:BTX9}) is evaluated numerically
with the prescribed $i\epsilon$ prescription at the light-cone
singularities. Together with the local response
$P_T(\alpha,\theta)$ derived above, this expression provides
the complete set of one-dimensional integrals required for
the numerical investigation of entanglement harvesting in
thermal field states.

An important consistency check is provided by the zero-temperature
limit. As $\theta\rightarrow0$, the hyperbolic cotangent terms
reduce to their inverse-argument limits, and Eq.~(\ref{eq:BTX9})
becomes

\begin{equation}
\begin{aligned}
\lim_{\theta\rightarrow0}
X_T(\alpha,\beta,\theta)
={}&
\frac{\lambda^2}{2\pi^{3/2}}
e^{-\alpha^2}
\\
&\times
\operatorname{PV}
\int_{-\infty}^{\infty}dx\,
\frac{e^{-x^2}}{4x^2-\beta^2}.
\end{aligned}
\label{eq:BTX10}
\end{equation}

Using the analytical evaluation of the principal-value integral
in Eq.~(\ref{eq:B12}), we recover the Minkowski vacuum result,

\begin{equation}
\lim_{\theta\rightarrow0}
X_T(\alpha,\beta,\theta)
=
-\frac{\lambda^2}{2\pi}
\frac{e^{-\alpha^2}}{\beta}
D\!\left(\frac{\beta}{2}\right)
=
X(\alpha,\beta).
\label{eq:BTX11}
\end{equation}

Thus, the thermal exchange amplitude continuously reduces to
its Minkowski vacuum counterpart in the zero-temperature limit.


\section{Uniformly Accelerated Detectors}
\label{app:Acceleration}

In this appendix we derive the integral representations used for the
uniformly accelerated detector calculations presented in the main text.
The local detector responses admit the same one-dimensional reduction
as in the thermal case, while the nonlocal exchange amplitude retains a
two-dimensional form because the two Born-rigid detector trajectories
have different proper-time parametrizations. The resulting expressions
form the basis of the numerical calculations presented in the main text.

For each detector, the local response is determined by the pullback of
the Minkowski-vacuum Wightman function to its uniformly accelerated
trajectory. This pullback has the form of a thermal correlation function
at the corresponding Unruh temperature. However, the two detectors have
different proper accelerations and hence different associated Unruh
temperatures. Their local excitation probabilities must therefore be
evaluated separately. The exchange amplitude, in contrast, depends on
both accelerated trajectories and must be evaluated using the
corresponding two-detector correlation function.

\subsection{Uniformly Accelerated Detector Trajectories and the
Wightman Function}

To investigate entanglement harvesting in uniformly accelerated frames,
we consider two Unruh--DeWitt detectors undergoing Born-rigid motion
along the $x$-direction. The detectors remain at fixed spatial
positions in the same Rindler frame while following hyperbolic
trajectories in Minkowski spacetime.

The transformation between Minkowski coordinates $(t,x)$ and Rindler
coordinates $(\eta,\xi)$ is
\begin{equation}
t=\rho\sinh(a\eta),
\qquad
x=\rho\cosh(a\eta),
\label{eq:C1}
\end{equation}
where
\begin{equation}
\rho=\frac{1}{a}+\xi.
\label{eq:C2}
\end{equation}

The detector trajectories correspond to constant Rindler spatial
coordinates. Detector $A$ is chosen to lie at
\[
\xi_A=0,
\]
while detector $B$ is located at
\[
\xi_B=L_R,
\]
where $L_R$ denotes the constant Rindler separation.

The corresponding worldlines are
\begin{equation}
\begin{aligned}
t_A(\tau_A)
&=
\frac{1}{a}\sinh(a\tau_A),
\\
x_A(\tau_A)
&=
\frac{1}{a}\cosh(a\tau_A),
\end{aligned}
\label{eq:C3}
\end{equation}
and
\begin{equation}
\begin{aligned}
t_B(\tau_B)
&=
\left(\frac{1}{a}+L_R\right)
\sinh\left(\frac{a\tau_B}{1+aL_R}\right),
\\
x_B(\tau_B)
&=
\left(\frac{1}{a}+L_R\right)
\cosh\left(\frac{a\tau_B}{1+aL_R}\right).
\end{aligned}
\label{eq:C4}
\end{equation}

The corresponding proper accelerations are
\begin{equation}
a_A=a,
\qquad
a_B=\frac{a}{1+aL_R}.
\label{eq:C4c}
\end{equation}
Although the detectors have different proper accelerations, they remain
at fixed Rindler spatial coordinates and maintain a constant proper
separation.

The natural evolution parameter of the Rindler frame is the common
Rindler time $\eta$, which coincides with the proper time of detector
$A$:
\begin{equation}
\tau_A=\eta.
\label{eq:C4d}
\end{equation}
For detector $B$, the proper time is related to the Rindler time by
\begin{equation}
\tau_B=(1+aL_R)\eta,
\label{eq:C4a}
\end{equation}
so that
\begin{equation}
d\tau_B=(1+aL_R)\,d\eta.
\label{eq:C4b}
\end{equation}
Thus, the two detectors share the same Rindler time coordinate, but
their proper times advance at different rates. This distinction is
important when evaluating the detector interaction Hamiltonian, whose
phase factors are expressed in terms of the respective proper times.

The invariant interval between events on the two trajectories, with
Rindler times $\eta$ and $\eta'$, is
\begin{equation}
(\Delta t)^2-(\Delta x)^2
=
-\frac{1}{a^2}
-\left(\frac{1}{a}+L_R\right)^2
+2\frac{1}{a}\left(\frac{1}{a}+L_R\right)
\cosh\left[a(\eta-\eta')\right].
\label{eq:C5}
\end{equation}
Using
\[
\cosh z=1+2\sinh^2\left(\frac{z}{2}\right),
\]
this becomes
\begin{equation}
(\Delta t)^2-(\Delta x)^2
=
\frac{4}{a}\left(\frac{1}{a}+L_R\right)
\sinh^2\left[\frac{a(\eta-\eta')}{2}\right]
-L_R^2.
\label{eq:C6}
\end{equation}

For a massless scalar field in four-dimensional Minkowski spacetime,
the Wightman function is
\begin{equation}
W(x,x')
=
-\frac{1}{4\pi^2}
\frac{1}{
(t-t'-i\epsilon)^2-(x-x')^2
},
\label{eq:C7}
\end{equation}
where the $i\epsilon$ prescription is understood in the usual
distributional sense.

For the local response of detector $A$, both points lie on the same
worldline. Taking the pullback of the Wightman function to this
trajectory gives
\begin{equation}
W_A(\tau_A,\tau_A')
=
-\frac{a^2}{16\pi^2}
\frac{1}{
\sinh^2\left[
\frac{a}{2}(\tau_A-\tau_A'-i\epsilon)
\right]
}.
\label{eq:C9}
\end{equation}
Similarly, the pullback to detector $B$ is
\begin{equation}
W_B(\tau_B,\tau_B')
=
-\frac{a_B^2}{16\pi^2}
\frac{1}{
\sinh^2\left[
\frac{a_B}{2}(\tau_B-\tau_B'-i\epsilon)
\right]
}.
\label{eq:C9b}
\end{equation}

Each local Wightman function has the thermal form associated with its
own Unruh temperature:
\begin{equation}
T_A=\frac{a}{2\pi},
\qquad
T_B=\frac{a_B}{2\pi}
=\frac{a}{2\pi(1+aL_R)}.
\label{eq:C10}
\end{equation}
Consequently, the local response of each detector can be related to
the response of an inertial detector in a thermal bath at the
corresponding Unruh temperature, provided the detector gap and
switching function are defined with respect to that detector's proper
time. This local correspondence does not imply that the two-detector
exchange amplitude is equivalent to that of two inertial detectors in
a common thermal bath.


\subsection{Local Detector Responses}

The leading-order excitation probability for detector $A$ is
\begin{equation}
P_A
=
\lambda^2
\int_{-\infty}^{\infty}d\tau_A
\int_{-\infty}^{\infty}d\tau_A'
\,
\chi_A(\tau_A)
\chi_A(\tau_A')
e^{-i\Omega(\tau_A-\tau_A')}
W_A(\tau_A,\tau_A').
\label{eq:C11}
\end{equation}
We take the switching function of each detector to be Gaussian in its
own proper time,
\begin{equation}
\chi_D(\tau_D)
=
\exp\left(-\frac{\tau_D^2}{2\sigma^2}\right),
\qquad D=A,B.
\label{eq:C11a}
\end{equation}

For detector $A$, introduce the relative and average proper-time
variables
\[
u=\tau_A-\tau_A',
\qquad
v=\frac{\tau_A+\tau_A'}{2}.
\]
The product of switching functions becomes
\[
\chi_A(\tau_A)\chi_A(\tau_A')
=
\exp\left(
-\frac{v^2}{\sigma^2}
-\frac{u^2}{4\sigma^2}
\right),
\]
and the integration over the average time gives
\[
\int_{-\infty}^{\infty}e^{-v^2/\sigma^2}\,dv
=
\sqrt{\pi}\sigma.
\]
Using Eq.~(\ref{eq:C9}), the response is therefore
\begin{equation}
P_A
=
-\frac{\lambda^2\sigma a^2}{16\pi^{3/2}}
\int_{-\infty}^{\infty}du\,
\frac{
e^{-u^2/(4\sigma^2)}e^{-i\Omega u}
}{
\sinh^2\left[\frac{a}{2}(u-i\epsilon)\right]
}.
\label{eq:C12}
\end{equation}

Introducing the dimensionless parameters
\begin{equation}
\alpha=\Omega\sigma,
\qquad
\gamma=a\sigma,
\qquad
\beta=\frac{L_R}{\sigma},
\label{eq:C13}
\end{equation}
and setting $x=u/(2\sigma)$, we obtain
\begin{equation}
P_A
=
-\frac{\lambda^2\gamma^2}{8\pi^{3/2}}
\int_{-\infty}^{\infty}dx\,
\frac{
e^{-x^2}e^{-2i\alpha x}
}{
\sinh^2\left[\gamma(x-i\epsilon)\right]
}.
\label{eq:C14}
\end{equation}
Here and below, the dimensionless regulator is understood to be
positive and to approach zero after the integral has been defined.

Define $P_T(\alpha,T\sigma)$ as the local excitation probability for an
inertial detector with gap $\Omega$, Gaussian switching width $\sigma$,
and temperature $T$. Since
\[
T_A\sigma=\frac{a\sigma}{2\pi}=\frac{\gamma}{2\pi},
\]
Eq.~(\ref{eq:C14}) gives
\begin{equation}
\boxed{
P_A(\alpha,\gamma)
=
P_T\left(\alpha,\frac{\gamma}{2\pi}\right).
}
\label{eq:C15}
\end{equation}

The response of detector $B$ must be evaluated separately. Its
excitation probability is
\begin{equation}
P_B
=
\lambda^2
\int_{-\infty}^{\infty}d\tau_B
\int_{-\infty}^{\infty}d\tau_B'
\,
\chi_B(\tau_B)
\chi_B(\tau_B')
e^{-i\Omega(\tau_B-\tau_B')}
W_B(\tau_B,\tau_B').
\label{eq:C16}
\end{equation}
Because the switching function is Gaussian in $\tau_B$, the same
change of variables used for detector $A$ yields
\begin{equation}
P_B
=
-\frac{\lambda^2\sigma a_B^2}{16\pi^{3/2}}
\int_{-\infty}^{\infty}du\,
\frac{
e^{-u^2/(4\sigma^2)}e^{-i\Omega u}
}{
\sinh^2\left[\frac{a_B}{2}(u-i\epsilon)\right]
}.
\label{eq:C17}
\end{equation}

Define
\begin{equation}
\gamma_B=a_B\sigma
=\frac{\gamma}{1+\gamma\beta}.
\label{eq:C18}
\end{equation}
Rescaling the integration variable as before gives
\begin{equation}
P_B
=
-\frac{\lambda^2\gamma_B^2}{8\pi^{3/2}}
\int_{-\infty}^{\infty}dx\,
\frac{
e^{-x^2}e^{-2i\alpha x}
}{
\sinh^2\left[\gamma_B(x-i\epsilon)\right]
}.
\label{eq:C19}
\end{equation}
The associated dimensionless Unruh temperature is
\[
T_B\sigma
=\frac{a_B\sigma}{2\pi}
=\frac{\gamma_B}{2\pi}.
\]
Consequently,
\begin{equation}
\boxed{
P_B(\alpha,\gamma,\beta)
=
P_T\left(
\alpha,
\frac{\gamma}{2\pi(1+\gamma\beta)}
\right).
}
\label{eq:C20}
\end{equation}

Thus, for Gaussian switching functions of the same width $\sigma$ in
the respective proper times, the two local responses are determined
by different dimensionless acceleration parameters:
\begin{equation}
\boxed{
P_A=P_T\left(\alpha,\frac{\gamma}{2\pi}\right),
\qquad
P_B=P_T\left(
\alpha,\frac{\gamma}{2\pi(1+\gamma\beta)}
\right).
}
\label{eq:C21}
\end{equation}
In general, these responses need not be equal. Their equality should
not be assumed solely because the detectors are stationary in the
same Rindler frame.

The formulas above use switching functions of equal width in the
detectors' respective proper times. If instead the switching protocol
is prescribed as a common function of Rindler time, the switching
function for detector $B$ must be transformed accordingly, and its
response must be recalculated with that convention.



\subsection{Exchange Amplitude}

We now consider the nonlocal exchange amplitude responsible for
entanglement harvesting between the two uniformly accelerated detectors.
Unlike the local response, the exchange amplitude probes field
correlations between two distinct worldlines and retains the geometric
information associated with the Born-rigid configuration.

Using the common Rindler time coordinate, the detector proper times
satisfy
\begin{equation}
\tau_A=\eta,
\qquad
\tau_B=(1+aL_R)\eta,
\label{eq:C22}
\end{equation}
so that
\begin{equation}
d\tau_A=d\eta,
\qquad
d\tau_B=(1+aL_R)\,d\eta.
\label{eq:C23}
\end{equation}
The switching functions are Gaussian in the detectors' respective
proper times, with the same proper-time width $\sigma$:
\begin{equation}
\chi_A(\eta)
=\exp\left(-\frac{\eta^2}{2\sigma^2}\right),
\qquad
\chi_B(\eta')
=\exp\left[-\frac{(1+aL_R)^2{\eta'}^2}{2\sigma^2}\right].
\label{eq:C24}
\end{equation}

At leading order in the detector-field coupling, the exchange amplitude
is
\begin{equation}
\begin{aligned}
X_A={}&-\lambda^2(1+aL_R)
\int_{-\infty}^{\infty}d\eta
\int_{-\infty}^{\infty}d\eta'\,
\chi_A(\eta)\chi_B(\eta')\\
&\quad\times e^{i\Omega\eta}
 e^{i\Omega(1+aL_R)\eta'}
 W\bigl(x_A(\eta),x_B(\eta')\bigr),
\end{aligned}
\label{eq:C25}
\end{equation}
where the Wightman function is evaluated between the two distinct
worldlines. In particular, its regulator is inherited from the
Minkowski-vacuum expression
\begin{equation}
W(x,x')=-\frac{1}{4\pi^2}
\frac{1}{(t-t'-i\epsilon)^2-(x-x')^2},
\label{eq:C26}
\end{equation}
with the $i\epsilon$ prescription taken in the distributional sense.

Introduce the dimensionless variables
\begin{equation}
u=\frac{\eta}{\sigma},
\qquad
v=\frac{\eta'}{\sigma},
\qquad
\alpha=\Omega\sigma,
\qquad
\beta=\frac{L_R}{\sigma},
\qquad
\gamma=a\sigma,
\label{eq:C27}
\end{equation}
so that $aL_R=\gamma\beta$. In these variables, the dimensionless
Minkowski coordinate differences between the two events are
\begin{align}
\delta t(u,v)
&=\frac{1}{\gamma}\sinh(\gamma u)
 -\frac{1+\gamma\beta}{\gamma}\sinh(\gamma v),
\label{eq:C28a}\\
\delta x(u,v)
&=\frac{1}{\gamma}\cosh(\gamma u)
 -\frac{1+\gamma\beta}{\gamma}\cosh(\gamma v).
\label{eq:C28b}
\end{align}
Define the dimensionless regulated denominator
\begin{equation}
\mathcal D_\epsilon(u,v)
=\bigl[\delta t(u,v)-i\epsilon\bigr]^2
 -\bigl[\delta x(u,v)\bigr]^2,
\label{eq:C29}
\end{equation}
where, in this equation, $\epsilon$ is measured in units of $\sigma$.
The exchange amplitude can then be written as the dimensionless
integral
\begin{equation}
\boxed{
\begin{aligned}
X_A={}&\frac{\lambda^2(1+\gamma\beta)}{4\pi^2}
\int_{-\infty}^{\infty}du
\int_{-\infty}^{\infty}dv\,
\frac{
\exp\left[-\frac{u^2}{2}
-\frac{(1+\gamma\beta)^2v^2}{2}\right]
 e^{i\alpha u}
 e^{i\alpha(1+\gamma\beta)v}
}{\mathcal D_\epsilon(u,v)}.
\end{aligned}
}
\label{eq:C30}
\end{equation}
This form keeps the Minkowski $i\epsilon$ prescription explicit. Away
from the regulator, the real part of the denominator is equivalent to
the invariant interval
\begin{equation}
(\Delta t)^2-(\Delta x)^2
=\sigma^2\left[
\frac{4(1+\gamma\beta)}{\gamma^2}
\sinh^2\left(\frac{\gamma(u-v)}{2}\right)
-\beta^2\right].
\label{eq:C31}
\end{equation}
The regulated denominator should, however, be formed from the Minkowski
coordinate differences as in Eq.~(\ref{eq:C29}), rather than by shifting
$u-v$ inside the simplified invariant-interval expression without
checking the regulator transformation.

Equation~(\ref{eq:C30}) depends only on the dimensionless parameters
$(\alpha,\beta,\gamma)$ and is evaluated as a two-dimensional
oscillatory integral. The two different proper-time parametrizations
prevent the exchange term from being replaced, in general, by a thermal
exchange amplitude evaluated at either detector's Unruh temperature.


\subsection{Negativity for Unequal Local Responses}

For completeness, we state the leading-order harvesting criterion when
the local excitation probabilities are not assumed to be equal. For
two initially unexcited detectors with identical coupling strength, the
leading-order negativity is
\begin{equation}
\boxed{
\mathcal N=\max\left\{0,
\frac{\sqrt{(P_A-P_B)^2+4|X_A|^2}-(P_A+P_B)}{2}
\right\}.
}
\label{eq:C32}
\end{equation}
Thus, entanglement harvesting occurs at this order when
\begin{equation}
\boxed{|X_A|^2>P_AP_B.}
\label{eq:C33}
\end{equation}
When $P_A=P_B=P$, Eq.~(\ref{eq:C32}) reduces to the familiar expression
$\mathcal N=\max(0,|X_A|-P)$. Since the two Born-rigid detectors have
different proper accelerations, equality of their local responses must
not be assumed without an independent justification. The local
responses in Eqs.~(\ref{eq:C15}) and (\ref{eq:C20}) should therefore be
used separately unless the switching protocol or a limiting case
establishes $P_A=P_B$.


\subsection{Numerical Evaluation}

The local responses are evaluated using the one-dimensional integrals
in Eqs.~(\ref{eq:C14}) and (\ref{eq:C19}), or equivalently through the
thermal-response function defined in Appendix~B. The relevant
dimensionless Unruh temperatures are
\begin{equation}
T_A\sigma=\frac{\gamma}{2\pi},
\qquad
T_B\sigma=\frac{\gamma}{2\pi(1+\gamma\beta)}.
\label{eq:C34}
\end{equation}
Accordingly, the numerical calculation must evaluate $P_A$ and $P_B$
separately, using the same detector gap and the same Gaussian width in
each detector's own proper time.

The exchange amplitude is evaluated from the two-dimensional integral
in Eq.~(\ref{eq:C30}). The Gaussian switching functions suppress the
integrand away from the origin, allowing the integrations to be
restricted to a finite computational domain, for example
$|u|,|v|\leq 8$, provided convergence is checked by enlarging the domain
and varying the numerical tolerance. The real and imaginary parts of
the integrand should be integrated separately before reconstructing the
complex exchange amplitude. Because the regulated denominator can be
sharply peaked near null-related points, the numerical prescription
must be checked for stability with respect to the regulator and
integration method.

For each point in the parameter space $(\alpha,\beta,\gamma)$, the
calculation should determine $P_A$, $P_B$, and $X_A$, and then evaluate
the reduced detector state and the negativity using
Eq.~(\ref{eq:C32}). The equal-response expression
$\mathcal N=\max(0,|X_A|-P_A)$ is appropriate only if the calculation
independently establishes $P_A=P_B$.

The accelerated-detector calculation therefore uses the same general
perturbative framework as the other detector configurations, but the
local responses and the nonlocal exchange amplitude must each be
computed with the trajectories, proper-time switching functions, and
Wightman-function prescription appropriate to the Born-rigid setup.

\bibliography{References.bib}
\end{document}